\documentclass[12pt]{article}
\usepackage{putex}
\usepackage{graphicx}
\usepackage{dsfont}
\usepackage{fancyvrb}

\newcommand{\PHI}{\Phi}
\newcommand{\gammat}{\gamma^{ t}}

\newcommand{\IRarrow}{\xrightarrow[\rm{IR}]{}}
\newcommand{\UVarrow}{\xrightarrow[\rm{UV}]{}}
\def\gYM{g_{\rm YM}}
\def\e{{\rm e}}

\newcommand{\psip}{\psi_+}
\newcommand{\psim}{\psi_-}

\begin{document}

\preprint{PUPT-2329}

\institution{PU}{Joseph Henry Laboratories, Princeton University, Princeton, NJ 08544, USA}

\title{Fermion correlators in non-abelian holographic superconductors}

\authors{Steven S. Gubser,\footnote{e-mail: {\tt ssgubser@Princeton.EDU}}
Fabio D. Rocha,\footnote{e-mail: {\tt frocha@Princeton.EDU}} and
Amos Yarom\footnote{e-mail: {\tt ayarom@Princeton.EDU}}}

\abstract{We consider fermion correlators in non-abelian holographic superconductors. The spectral function of the fermions exhibits several interesting features such as support in displaced Dirac cones and an asymmetric distribution of normal modes. These features are compared to similar ones observed in angle resolved photoemission experiments on high $T_c$ superconductors. Along the way we elucidate some properties of $p$-wave superconductors in AdS${}_4$ and discuss the construction of $SO(4)$ superconductors.}

\date{February 2010}

\maketitle

\tableofcontents

\section{Introduction}
\label{INTRO}

Holographic superconductors are curved geometries which exhibit spontaneous breaking of a gauge symmetry and are dual to field theories which correspondingly exhibit spontaneous breaking of a global symmetry. 
The dual field theories are typically large $N$ gauge theories with relativistic conformal invariance.  The symmetry breaking order parameter is the expectation value of a composite operator in the field theory which is a singlet under all the gauge symmetries of the field theory, but transforms under the global symmetries that are gauged in the bulk.\footnote{The global symmetry of the field theory which the order parameter breaks can in principle be weakly gauged.  This is analogous to BCS theory, where phonon-electron interactions are the crucial dynamics of symmetry breaking, and the $U(1)$ of electromagnetism can be treated as a global symmetry for most purposes: For example, conductivity is computed using a two-point function of the Noether current corresponding to the $U(1)$ symmetry before it is gauged.}  
In holographic superconductors, the broken symmetry phase has a finite charge density. This distinguishes holographic superconductors from vacuum configurations with spontaneous symmetry breaking. Holographic superconductors whose order parameter is a scalar under Lorentz transformations are called $s$-wave superconductors. These were first discussed in \cite{Gubser:2008px,Hartnoll:2008vx}.  Holographic superconductors with a vector order parameter are called $p$-wave superconductors.  Early studies include \cite{Gubser:2008zu,Gubser:2008wv,Roberts:2008ns}.

Consider as an example a dual field theory which is four-dimensional and includes chiral fermions $\lambda_\alpha$ and $\bar\kappa^{\dot\alpha}$, both transforming in the complexified adjoint of an $SU(N)$ gauge group, and both carrying charge $-e$ under the $U(1)$ symmetry that gets spontaneously broken.  Then the symmetry breaking order parameter could be $\langle \tr \lambda^\alpha \lambda_\alpha \rangle$ in the case of an $s$-wave holographic superconductor, or $\langle \tr \lambda^\alpha \sigma^1_{\alpha\dot\beta} \bar\kappa^{\dot\beta} \rangle$ in the case of a $p$-wave superconductor, where $\sigma^1$ is the first Pauli matrix.  There is no lattice structure in the dual field theory; thus the breaking of rotational invariance by the $p$-wave condensate is spontaneous. The holographic treatment of both the $s$-wave and $p$-wave cases is typically at the level of mean field theory in that one solves classical equations of motion in the bulk without inquiring about the role of fluctuations.  This is justified on the field theory side if one is restricting attention to leading order effects in a large $N$ expansion, where $N$ is the rank of the gauge group.\footnote{In \cite{Franco:2009yz,Aprile:2009ai,Franco:2009if} non mean field theory behavior was exhibited by arranging for a particular bulk lagrangian for the scalar field.}  Notable reviews of holographic superconductors include \cite{Hartnoll:2009sz,Herzog:2009xv,Horowitz:2010gk}.

One useful probe of the electronic structure of high-$T_c$ superconductors is the angle resolved photoemission experiments (ARPES) which essentially rely on the photoelectric effect and measure the energy of the electrons emitted from the sample. ARPES experiments demonstrate some interesting properties of high $T_c$ superconductors.  These properties include a gap with $d_{x^2-y^2}$ symmetry, deformed Dirac cones whose apexes are the nodes of the superconducting gap, Fermi arcs whose zero temperature limits are the nodes in the gap, and a peak-dip-hump structure of the emission intensity as a function of frequency at fixed wave-number.  A review of ARPES measurements can be found in \cite{RevModPhys.75.473}.

Holographic superconductors differ from high-$T_c$ superconductors in several respects: Most notably, they are symmetry-breaking states of large $N$ gauge theories, and they have no underlying lattice structure. Nevertheless, it is interesting to ask whether or not the properties of the fermionic spectral function in high $T_c$ superconductors are shared by their holographic counterparts. In \cite{Chen:2009pt,Faulkner:2009am,Gubser:2009dt}, following \cite{Lee:2008xf,Cubrovic:2009ye,Liu:2009dm}, an analysis of fermion correlation functions was carried out for the holographic $s$-wave superconductors introduced in \cite{Gubser:2008px,Hartnoll:2008vx}. In this work, after elucidating some details of the phase diagram of the holographic $p$-wave superconductors in AdS${}_4$, we discuss some properties of its fermionic correlation functions.  Along the way we introduce a variant of the holographic $p$-wave construction, based on an $SO(4)$ gauge group.

This work is organized as follows. In section \ref{PWAVE} we review the construction of a $p$-wave superconductor in AdS${}_4$ and study its phase diagram. An interesting property of the $p$-wave superconductor, first observed in \cite{Basu:2009vv}, is that the bulk geometry is a domain wall interpolating between infrared and ultraviolet limits which are each AdS${}_4$ with an asymptotically flat gauge connection.\footnote{To be precise, the gauge connection has a field strength whose stress tensor becomes insignificant near the boundary compared to the negative cosmological constant.  It is in this sense that the ultraviolet geometry can be approximated by AdS${}_4$ with a flat gauge connection.} In section \ref{CONFORMAL} we discuss in detail scalar, fermion, and vector correlation functions in the asymptotic regions of such backgrounds, starting from a general gauge group.  In section \ref{SPECTRAL} we explain how to compute the fermion spectral function in the full, zero-temperature, domain-wall, $p$-wave background and described some of its properties. The numerical results of the computation can be found in section \ref{FERMIONS}. In section \ref{DISCUSSION} we discuss the results in the context of ARPES experiments on high $T_c$ superconductors. The reader familiar with the $p$-wave superconductor background, and interested only in the features of the spectral function should have a look at sections \ref{SUMMARY}, \ref{FERMIONS} and \ref{DISCUSSION}. The reader unfamiliar with computations of correlators in the context of AdS/CFT may find section \ref{CONFORMAL} useful.

\section{The $p$-wave holographic superconductor}
\label{PWAVE}

The simplest example of a non-abelian holographic superconductor is the $p$-wave superconductor first introduced in \cite{Gubser:2008wv} (following earlier work \cite{Gubser:2008zu}) and further studied in \cite{Roberts:2008ns,Basu:2009vv} and \cite{Basu:2008bh,Herzog:2009ci,Ammon:2009xh,Zeng:2009dr}.
The bulk action for the $p$-wave superconductor is given by
\begin{equation}
\label{E:PwaveAction}
  S = \int_M d^4 x \, \sqrt{-g} \, \left(
  R +\frac{6}{L^2}  - {1 \over 2} \tr F_{\mu\nu}^2 \right) \,.
\end{equation}
In what follows we will set $L=1$. 
Here $F_{\mu\nu}$ is the field strength of an $SU(2)$ gauge potential:
 \eqn{FDef}{
  F_{\mu\nu} = \partial_\mu A_\nu - \partial_\nu A_\mu - i \gYM [A_\mu,A_\nu] \,,
 }
and
\begin{equation}
	A_{\mu}  = A^{a}_{\mu} \tau^{a}\,,
\end{equation}
and $\tau^a = {1 \over 2} \sigma^a$, where $\sigma^a$ are
the Pauli matrices. Consider a configuration in which the gauge field takes the form:
\begin{equation}
\label{AforSU2}
	A = \Phi(r)\tau^3 dt + W(r) \tau^1 dx.
\end{equation}
Based on the general arguments in \cite{Gubser:2008zu,Gubser:2008wv}, one expects that apart from the Reissner-Nordstrom solution for which $W=0$, there exist other solutions where $W\neq0$, corresponding to a non-zero expectation value of the boundary current $J^1_x$.  This expectation value spontaneously breaks both the $SU(2)$ gauge symmetry and rotational invariance.
Symmetry breaking solutions of this type have been studied in the limit where $\gYM \to \infty$ in \cite{Gubser:2008wv} and in a conjectured zero temperature configuration in \cite{Basu:2009vv}. In \cite{Ammon:2009xh}, five dimensional, non-zero temperature and finite gauge coupling geometries were studied. The purpose of this section is to fill a gap in the literature by studying the AdS${}_4$ $p$-wave superconductor geometry for finite temperature and coupling.  Through this numerical study we will confirm that the domain wall geometries of \cite{Basu:2009vv} are indeed the zero-temperature limits of symmetry-breaking black holes.

Parametrizing the line element by 
\begin{equation}
\label{gforSU2}
	ds^2 = -r^2 \gamma(r)e^{-\chi(r)} dt^2 + \frac{dr^2}{r^2 \gamma(r)} + r^2(c(r)^2dx^2 + dy^2)\,,
\end{equation}
we find that the equations of motion for the gauge field and the metric are
\begin{align}
\label{E:BackgroundEOM}
\begin{split}
	\frac{1}{2} r W' \left(2 r \gamma ' -\frac{2 r c'  \gamma }{c}-r \chi '  \gamma +4  \gamma \right )+\frac{ \gYM^2 W   \Phi  ^2 e^{\chi  }}{r^2 \gamma  }+r^2 \gamma   W'' & = 0\\
	\frac{1}{2} r^2 \gamma    \Phi '  \left(\frac{2 c' }{c }+\chi ' +\frac{4}{r}\right)-\frac{ \gYM^2 W ^2  \Phi  }{r^2 c ^2}+r^2 \gamma    \Phi '' & = 0 \\
	\frac{c' }{c } \left(\frac{\gamma ' }{\gamma  }-\chi ' +\frac{2}{r}\right)-\frac{ \gYM^2 W ^2  \Phi  ^2 e^{\chi  }}{r^6 c ^2 \gamma  ^2}-\frac{\chi ' }{r}& = 0 \\
	-\frac{r^2 \gamma   c'' }{c }-\frac{r c'  \left(r \gamma ' +8 \gamma  \right)}{2 c }-\frac{ \gYM^2 W ^2  \Phi  ^2 e^{\chi  }}{4 r^4 c ^2 \gamma  }+\gamma     \left(-\frac{{W'}^2}{4 c ^2}-3\right)-r \gamma ' -\frac{1}{4} e^{\chi  } { \Phi '} ^2+3&=0 \\
	c  c'' +c'  c  \left(\frac{\gamma ' }{\gamma  }+\frac{4}{r}-\frac{1}{2}  \chi ' \right)-\frac{ \gYM^2 W ^2  \Phi  ^2 e^{\chi  }}{2 r^6 \gamma ^2}+\frac{{W'} ^2}{2 r^2}&=0\,.
\end{split}
\end{align}
We have omitted an additional equation of motion which is automatically satisfied once the gauge fields and metric components solve \eqref{E:BackgroundEOM}.  In the limit where $\gYM \to \infty$, the matter content of the theory decouples from gravity and the equations of motion reduce to gauge fields in an AdS${}_4$-Schwarzschild or AdS${}_4$ background. This is the probe limit, initially studied in \cite{Gubser:2008wv}, following earlier work \cite{Hartnoll:2008vx} on a similar limit of the holographic Abelian Higgs model.  In the following subsection we will solve the equations of motion numerically and discuss some of the features of the solution. We will revisit the probe limit in section \ref{PROBEBGD}.

\subsection{Numerics and phase diagram}
\label{NUMERICSBGD}

The equations for $W$, $\Phi$ and $c$ are second order while the equations for $\gamma$ and $\chi$ are first order. Thus, to obtain a solution, we need to specify eight integration constants. In the deep infrared (IR), located at $ r=0$, we set $\Phi \IRarrow 0$, $W \IRarrow {\rm finite}$, $\gamma \IRarrow {\rm finite}$ and $c \IRarrow {\rm finite}$. (If we are looking for finite temperature solutions then we should require that $\gamma$ vanishes at the horizon $r=r_{\rm H}$, and if we are looking for zero temperature solutions we should require that $\gamma$ is finite in the deep IR). Near the boundary, located at $r\to\infty$, we require that $W \UVarrow 0$ and that $\Phi \UVarrow \mu$, $\mu$ being the chemical potential of the boundary theory. The remaining two integration constants can be thought of as the values of $\chi$ and $c$ at the boundary, and these can be gauged to $0$ and $1$, respectively, by rescaling the $t$ and $x$ coordinates. In practice, we've looked for solutions by using a standard shooting algorithm from the horizon to the boundary.

\begin{figure}
\begin{center}
\includegraphics[width = 8 cm]{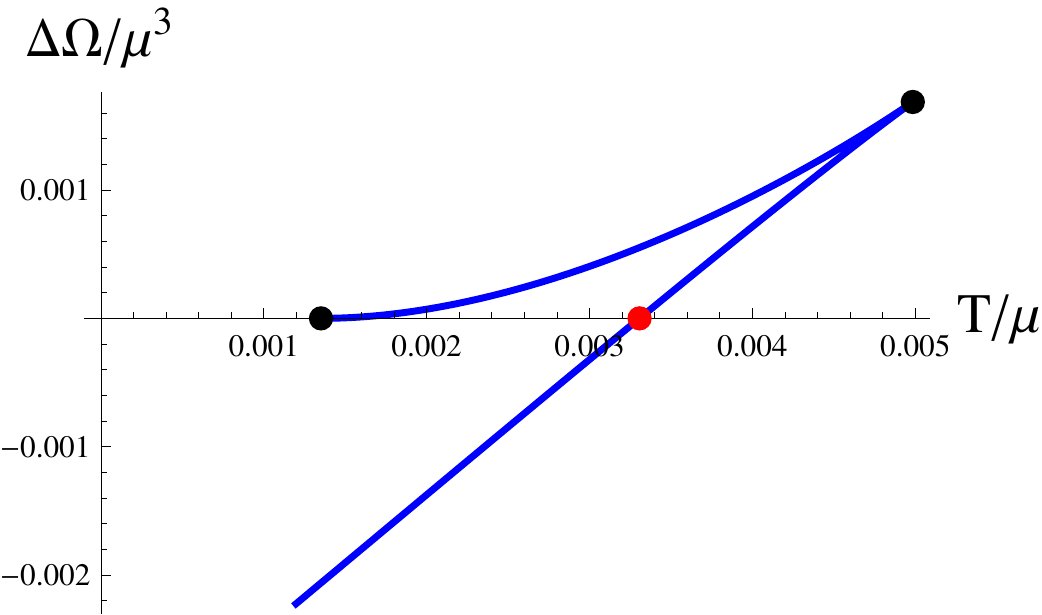}
\includegraphics[width = 8 cm]{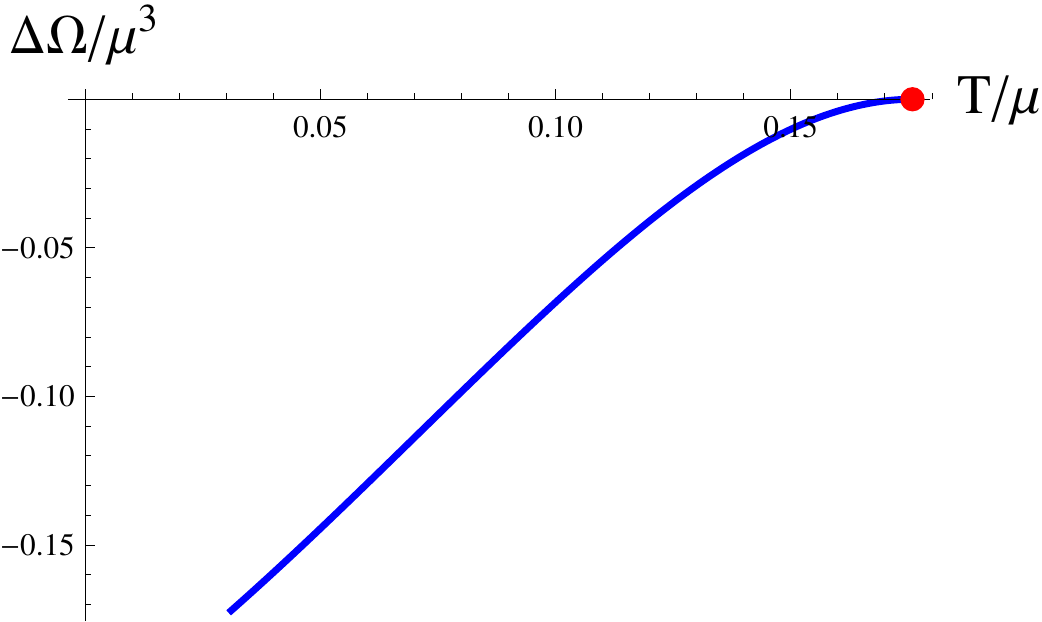}
\caption{\label{F:PTOmega} The difference between the grand canonical potential for the condensed phase and the uncondensed phase, $\Delta \Omega$. The red dots show the location of the critical temperature, and the black dots the location of the spinodal points. The left plot corresponds to $\gYM=0.79$ and the right one to $\gYM=3$.}
\end{center}
\end{figure}

To analyze the stability of the solutions with $W\neq 0$ we compute the boundary theory grand canonical potential $\Omega$ (per unit volume) of these configurations and compare it to the grand canonical potential of the Reissner-Nordstrom black hole. The computation of $\Omega$ is carried out by computing the on-shell Euclidean action $\Omega = -T S_{\rm E}/V$. We refer the reader to \cite{Ammon:2009xh} for the details of a similar computation in AdS${}_5$. We find that solutions with a non-vanishing condensate, $W\neq 0$, are stable only below a critical temperature $T_c$, which varies with the charge $\gYM$. Above $T_c$ the canonical potential for the AdS Reissner-Nordstrom black hole is lower, and it is the preferred solution (see figure \ref{F:PTOmega}). We find that the phase transition from the RN solution to the condensed solution is second order if $\gYM > 1.14 \pm 0.01$, and is first order if $\gYM < 1.14 \pm 0.01$. When $\gYM < 0.710 \pm 0.001$ the condensed solution no longer exists.\footnote{With 20 digits of working precision, the lowest value of $q$ for which Mathematica's NDSolve algorithm could obtain a solution was $q=0.7103$.}
\begin{figure}
\begin{center}
\includegraphics[width = 12 cm]{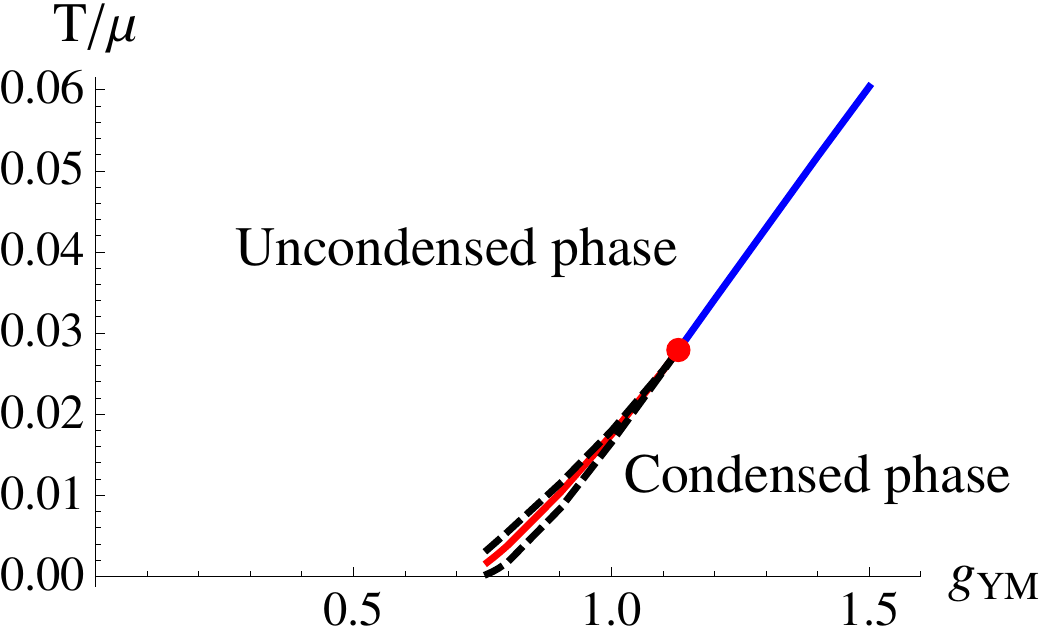}
\caption{\label{F:PTtotal} The phase diagram for the $p$-wave superconductor. The blue curve indicates a second order transition and the red line a first order phase transition. The dashed black lines are spinodal curves and the red dot marks a tricritical point.}
\end{center}
\end{figure}
\begin{figure}
\begin{center}
\includegraphics[width = 16 cm]{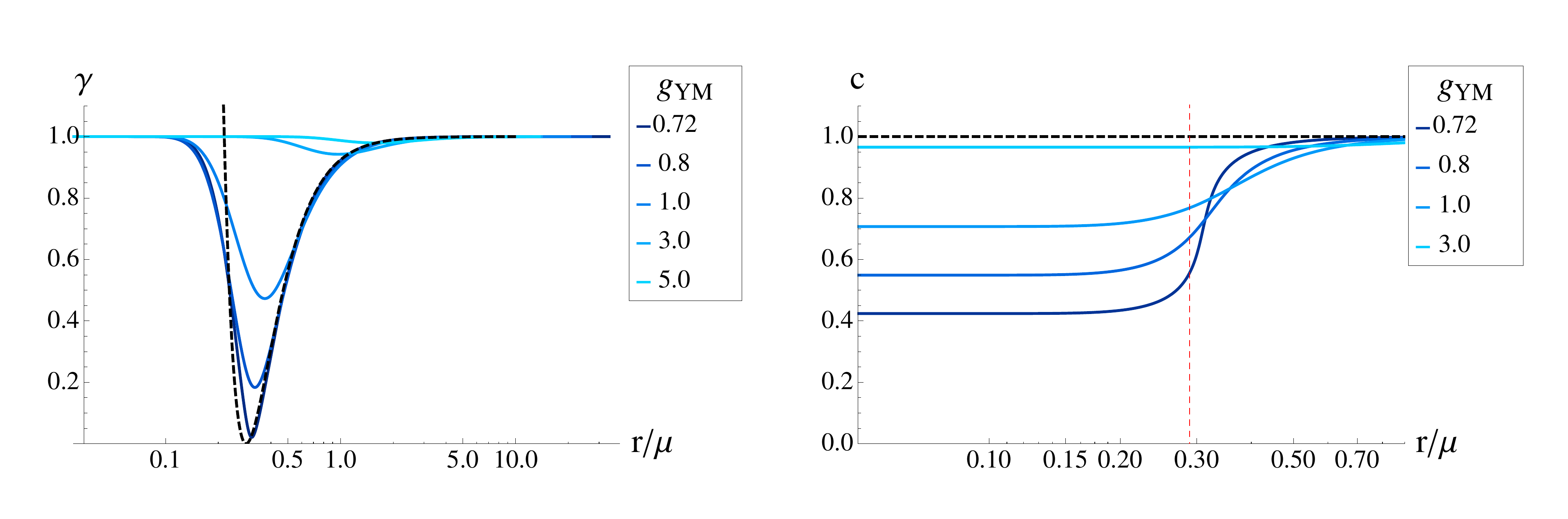}
\caption{\label{F:Approachqc} Plots of the metric components $\gamma$ and $c$ at zero temperature as a function of the the charge. As the charge approaches its critical value, the black hole hair gets pushed further into the IR, conforming to an extremal Reissner-Nordstrom black hole in the UV. The dashed black line corresponds to the extremal Reissner-Nordstrom solution. The dashed vertical red line on the right plot signifies the location of the extremal RN horizon where $\gamma = 0$.}
\end{center}
\end{figure}

In \cite{Basu:2009vv} it was conjectured that the zero temperature limit of the condensed phase is a domain wall geometry similar to the one described in \cite{Gubser:2008wz}: The infrared and ultraviolet geometries are both asymptotically AdS but with different speeds of light. The condensate $W$, the gauge field $\Phi$ and the metric components $\chi$ and $c$ interpolate between their infrared values $W_{\rm IR}>0$, $c_{\rm IR}>1$, $\chi_{\rm IR} \neq 0$ and $\Phi_{\rm IR}=0$ in the infrared to their UV values $W_{\rm UV} = 0$, $c_{\rm UV} = 1$, $\chi_{\rm UV} = 1$ and $\Phi_{\rm UV} = \mu>0$. $\gamma$ approaches 1 both in the UV and in the IR. Note that since $c$ differs in the UV and IR, the appropriate speeds of light differ in the $x$ and $y$ directions.  This anisotropy is a new and distinctive feature of the domain wall of \cite{Basu:2009vv}.  Another interesting feature is that the AdS radius is the same in the ultraviolet and infrared: indeed, both limits are just empty AdS${}_4$ with a flat $SU(2)$ gauge connection.

\begin{figure}
\begin{center}
\includegraphics[width = 12 cm]{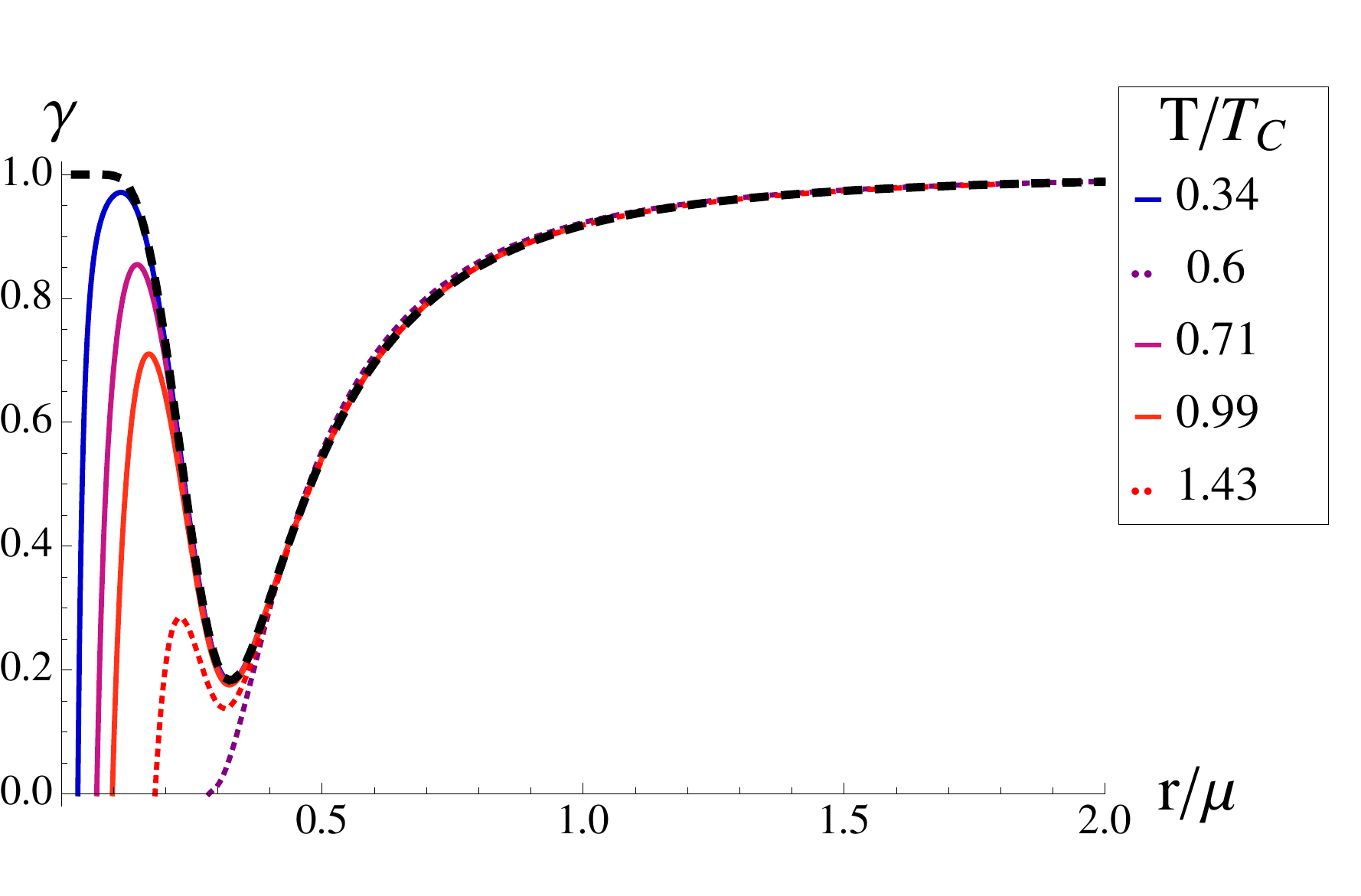}
\caption{\label{F:Approach0} Plots of $\gamma$ as a function of the temperature for $q=0.8$. The curves are color coded according to the temperature of the solution. The dashed line shows the domain wall solution and the dotted lines correspond to superheated or supercooled solutions.}
\end{center}
\end{figure}

By following the branch of symmetry-breaking solutions from the region where $W$ is perturbatively small to the region where it is larger, we were able to
verify that the domain wall geometries described in the previous paragraph are indeed the zero temperature limits of solutions with regular, finite-temperature horizons. See figure \ref{F:Approach0}.
From the phase diagram of the $SU(2)$ superconductor, depicted in figure \ref{F:Approachqc}, and by comparing the domain wall geometry of the condensed phase with subsequently small $\gYM$ to that of the extremal RN solution (figure \ref{F:Approachqc}), it appears that crossing $\gYM = 0.710 \pm 0.001$ at $T=0$ results in a phase transition from the condensed solution to the extremal RN solution.

\subsection{The probe limit}
\label{PROBEBGD}

As discussed earlier, in the probe limit the matter content of the theory decouples from the metric and the equations of motion \eno{E:BackgroundEOM} reduce to the Yang-Mills equation in a fixed background geometry.  For convenience we rescale the gauge field to get rid of all the factors of $\gYM$ in the equations of motion: that is, $\Phi \to \Phi/\gYM$ and $W \to W/\gYM$.  This has the same effect as setting $\gYM=1$ in the Yang-Mills equations---but we should bear in mind that the probe approximation is justified precisely by taking $\gYM$ large.  An additional simplification is possible at zero temperature: we recall that AdS${}_4$ is conformally flat, and that the classical Yang-Mills equations in four dimensions are conformally invariant.  Explicitly, if the metric is expressed as
 \eqn{AdSfourSimple}{
  ds^2 = {L^2 \over z^2} (-dt^2 + dx^2 + dy^2 + dz^2) \,,
 }
then systematically dropping the overall prefactor $L^2/z^2$ has no effect on the Yang-Mills equations.  With the ansatz \eno{AforSU2}, they take the form
 \eqn{ProbeEOMs}{
  \Phi'' = W^2 \Phi \qquad
  W'' = -\Phi^2 W \,,
 }
where primes denote $d/dz$.  The boundary conditions appropriate for describing the type of domain wall solution we are interested in are
 \eqn{BCs}{\seqalign{\span\TR & \quad\hbox{and}\quad \span\TR &\qquad\qquad\span\TT}{
  \Phi \to 0 & W \to W_{\rm IR} & at $z=0$  \cr
  \Phi \to \mu & W \to 0 & as $z \to +\infty$ \,,
 }}
where $W_{\rm IR}$ and $\mu$ are finite.  Even though one can find one conserved charge for this system \cite{Yarom:2009uq} (namely the Hamiltonian associated with radial translations), the equations of \eno{ProbeEOMs} do not appear to be integrable. A closely related system studied in \cite{Savvidy:1982wx} is known to exhibit strongly chaotic behavior.

\begin{figure}
  \centerline{\includegraphics[width=4in]{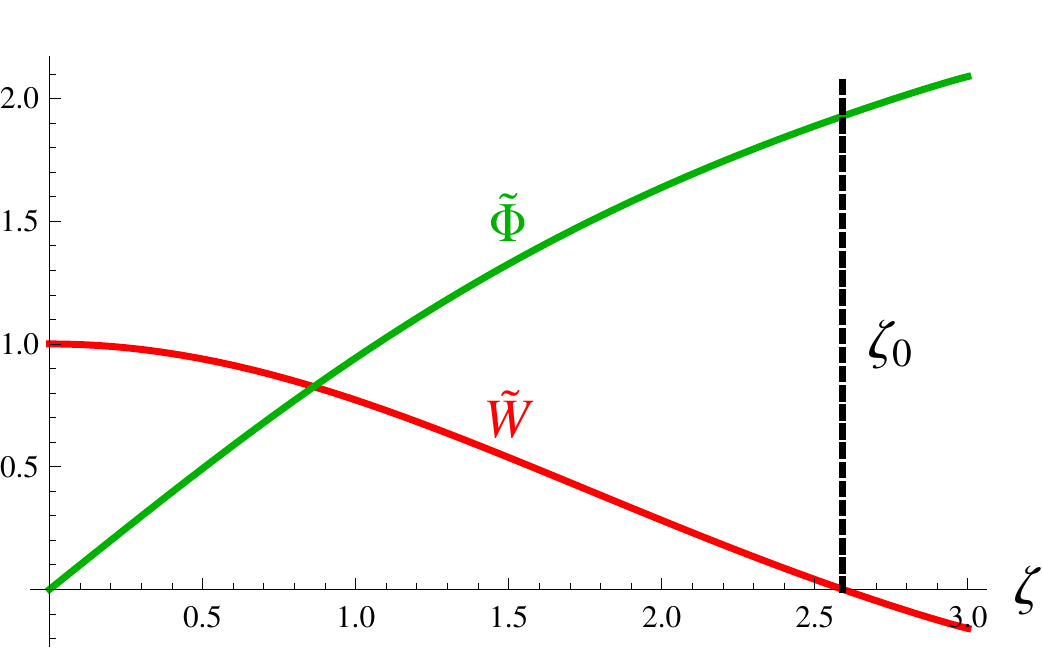}}
  \caption{The approximate functions $\tilde\Phi(\zeta)$ and $\tilde{W}(\zeta)$ evaluated from \eqref{E:Xseries} with $n=40$, with the positive root of $\tilde{W}$ shown.}\label{FIG:SeriesYM}
 \end{figure}

We find it convenient to change variables and make the field redefinitions as follows:
 \eqn{UniqueAnsatz}{
  \Phi(z) = W_{\rm IR} \tilde\Phi(\zeta) \qquad
  W(z) = W_{\rm IR} \tilde{W}(\zeta) \,,
 }
where
 \eqn{zetaDef}{
  \zeta = e^{-W_{\rm IR}z}\zeta_0 \,,
 }	
and $\zeta_0$ is a constant yet to be determined.  Large $z$ now corresponds to $\zeta = 0$, and we can reformulate the boundary value problem by requiring that at $\zeta=0$,
\begin{equation}
	\tilde{W}(0)=1\qquad \tilde{\Phi}(0) = 0 \qquad \tilde{\Phi}'(0)=1 \qquad \tilde{W}'(0)=0\,.
\end{equation}
The UV is located at $\zeta = \zeta_0$ and $\zeta_0$ is the smallest $\zeta$ for which
\begin{equation}
	\tilde{W}(\zeta) = 0.
\end{equation}
The equation for $X = \tilde{W} + i \tilde{\Phi}$ is
\begin{equation}
	\zeta^2 \partial_\zeta^2 X + \zeta \partial_\zeta X + \frac{1}{4} X ((X^*)^2- X^2) = 0
\end{equation}
which we can solve by Taylor expanding around $\zeta=0$. Defining
\begin{equation}
\label{E:Xseries}
	X = \sum_{n=0}^{\infty} i^n \alpha_n \zeta^n\,,
\end{equation}
we find that the $\alpha_n$'s satisfy:
\begin{equation}
\label{E:alphaval}
	n^2 \alpha_n = \sum_{k=0}^{\left[\frac{n-1}{2}\right]} \sum_{p=0}^k \alpha_{n-(2k+1)}\alpha_{2k+1-p} \alpha_p
\end{equation}
where $\left[(n-1)/2\right]$ means the largest integer that is smaller or equal to $(n-1)/2$.
Equation \eqref{E:alphaval} can be solved recursively once we are given $\alpha_0=1$ and $\alpha_1 = 1$. The first few terms in the expansion are
 \eqn{SeriesForm}{
  \tilde{W}(\zeta) + i \tilde\Phi(\zeta) &= 1 + i \zeta - {1 \over 4} \zeta^2 - 
    {i \over 16} \zeta^3 + {3 \over 128} \zeta^4 + {3i \over 512} \zeta^5 - 
    {1 \over 512} \zeta^6 - \ldots
}
and we note that the quantities $4^n \, n! \, n!! \, \alpha_n$ are positive integers at least to $n=100$.
Once again, by considering the first 100 terms of the series, we approximate the radius of convergence of the Taylor series expansions for $X$ to be $3.38$.  Between $0$ and $3.38$, $\tilde{W}$ has a zero (meaning the phase of $X$ is $i\pi/2$) that appears to be unique.  It is at
 \eqn{zetaStar}{
  \zeta_0 \approx 2.5918 \,.
 }
See Fig.~\ref{FIG:SeriesYM}.

One may now extract $\mu = \tilde{\Phi}(\zeta_0) W_{\rm IR} \approx 1.9285 W_{\rm IR} $.  Note that the range of $\zeta$ corresponding to $0 < z < \infty$ is $\zeta_0 > \zeta > 0$.  It is convenient that the radius of convergence of the Taylor series \eno{SeriesForm} is significantly larger than $\zeta_0$: because of this, we can get uniformly good accuracy for $\tilde\Phi$ and $\tilde{W}$ on the interval of interest from the Taylor series expansion.  In practice we terminated the series \eno{SeriesForm} at $n=100$. In section \ref{FERMIONS} we will approximate this $T=0$ background with a sharp domain wall solution 
depicted in figure \ref{FIG:SeriesYM}.

\section{Two-point functions from conformally flat backgrounds}
\label{CONFORMAL}

With zero-temperature and finite-temperature backgrounds in hand, an obvious next step is to compute Green's functions associated with interesting physical phenomena, such as conductivity and photoemission.  Such computations typically rely heavily on numerics, because even the coefficients in the differential equations to be solved are known only numerically.  It turns out, however, that some of the features of the final answer can be understood qualitatively in terms of the Green's functions one would get for the ultraviolet and infrared AdS${}_4$ geometries alone.  These asymptotic forms can be obtained analytically.  This section is devoted to a general study of such Green's functions for more general gauge groups and even for slightly more general gravitational backgrounds than AdS${}_4$.  The backgrounds we study are conformally flat, and we require that the gauge connection is also flat: that is, there is no Yang-Mills field strength.  
In the context of AdS/CFT, the flat Yang-Mills connection in the bulk implies that the field theory lagrangian on the boundary has been explicitly deformed by components of conserved currents associated with continuous symmetries, and that all the deformations commute both with each other and the original Hamiltonian.  As explained earlier, our motivation for carrying out this analysis is to get a better understanding of the infrared and ultraviolet behavior of the domain wall geometries of the non-abelian holographic superconductors described in the previous subsection.  Our analysis is, however, more general. We will see that we can understand not only the fermion correlators, but also the UV and IR limits of scalar and vector correlation functions for a generic non-abelian gauge group.

The calculations in this section, though slightly abstract when expressed in terms of an arbitrary semi-simple Lie algebra and arbitrary unitary representations thereof, are in essence formal elaborations of the simplest calculations possible in AdS/CFT.  We have even avoided the use of curved geometries by restricting attention to the conformally coupled scalar, the massless Dirac fermion, and non-abelian gauge fields with a Yang-Mills action.  The equations we solve can be understood as describing the free propagation of massless fields in flat space---with a few mild modifications dictated by the bulk gauge symmetries.  Thus, this section is relatively self-contained and may serve as a useful introduction to holographic Green's functions for formally inclined readers.

Readers wishing to pass over the technical details but understand the qualitative features may wish to skip straight to section~\ref{SUMMARY}.

\subsection{The action}
\label{CONFORMALACTION}

Consider the following action for gravity, a Yang-Mills field of a simple gauge group with Lie algebra ${\bf g}$, a complex scalar field $\Sigma$ in the unitary representation ${\bf r}_\Sigma$ of ${\bf g}$, a Dirac fermion $\Psi$ in the unitary representation ${\bf r}_{\Psi}$ of ${\bf g}$ and a real singlet scalar $\phi$:
 \eqn{SSUGRA}{
  S &= \int_M d^4 x \, \sqrt{-g} \, \Bigg[
    \left( 1 - {1 \over 6} |\Sigma|^2 \right) R - {1 \over 2} \tr F_{\mu\nu}^2 -
    i\bar\Psi \Gamma^\mu D_\mu \Psi -
     {1 \over 2} (\partial\phi)^2 - |D_\mu \Sigma|^2
      - V(\phi) \Bigg] + S_{\rm bdy} \,.
 }
Ordinarily there is a prefactor $1/2\kappa^2 = 1/16\pi G_N$ multiplying $S$, but we drop this factor.  We normalize the trace $\tr$ in \eno{SSUGRA} so that $\tr t^a t^b = {1 \over 2} \delta^{ab}$, where the $t^a$ are generators of ${\bf g}$. Since the gauge group is simple, the field strength is given by
 \eqn{FDefAgain}{
  F_{\mu\nu} = \partial_\mu A_\nu - \partial_\nu A_\mu - i \gYM [A_\mu,A_\nu] \,,
 }
and we have defined
 \eqn{DDef}{
  D_\mu \Sigma &= (\partial_\mu - i \gYM A^a t^a_{{\bf r}_\Sigma}) \Sigma  \cr
  D_\mu \Psi &= \left( \partial_\mu + {1 \over 4} \omega_\mu{}^{\underline{\rho\sigma}}
    \Gamma_{\underline{\rho\sigma}} - i \gYM A_\mu^a t^a_{{\bf r}_\Psi} \right) \Psi \,.
 }
Here $t^a_{\bf r}$ are the generators of ${\bf g}$ in the representation ${\bf r}$, $\omega_\mu{}^{\underline{\rho\sigma}}$ is the spin connection, and underlined Greek indices are tangent space indices.  When there is no risk of confusion, we will omit the subscript on representation matrices like $t^a_{{\bf r}_\Sigma}$: so, for example, $t^a \Sigma = t^a_{{\bf r}_\Sigma} \Sigma$.  We define gamma matrices so that $\Gamma^{\underline{0}}$ is anti-hermitian, while the other $\Gamma^{\underline\mu}$ are hermitian, and so that $\{ \Gamma^{\underline\mu}, \Gamma^{\underline\nu} \} = 2\eta^{\underline{\mu\nu}}$ where $\eta^{\underline{\mu\nu}} = \diag\{ -1,1,\ldots,1 \}$.  Also, we define $\bar\Psi = \Psi^\dagger \Gamma^{\underline{t}}$. The terms in the action $S_{\rm bdy}$ includes boundary terms which do not affect the equations of motion but render the on-shell action finite.

There are several simple ways in which we could generalize \eno{SSUGRA}---but typically will not for the purposes of this section.  First, we could generalize to gauge group that are semi-simple or that contain abelian factors.  The main difference is that then one has multiple independent gauge couplings.  Second, we could introduce an explicit mass term for $\Sigma$, and/or other $\Sigma$ dependence in $V$.  Third, we could introduce a Dirac mass for $\Psi$, and in the case of a real representation for fermions, also a Majorana mass.  And fourth, depending on the choice of representations ${\bf r}_\Sigma$ and ${\bf r}_\Psi$, we could introduce some Yukawa interactions.

The action \eno{SSUGRA} is similar to what one typically finds for the gauged supergravity truncations of the near-horizon dynamics of D- and M-brane constructions.  It was chosen for purposes of producing simple illustrations of the Green's function computations that we want to do.  The biggest difference between \eno{SSUGRA} and supergravity actions is that in supergravity, there are spin-$3/2$ fields (the gravitini) in addition to $g_{\mu\nu}$, $A_\mu$, $\Psi$, $\Sigma$, and $\phi$.  Further differences are that in gauged supergravity, one would typically redefine the metric through a scalar-dependent Weyl transformation so that the Einstein-Hilbert term has no prefactor; and that in supergravity, the kinetic terms of the other fields would typically involve interesting functions of the scalars.

\subsection{Conformally flat backgrounds}
\label{CONFORMALBG}

The simplest class of solutions to \eno{SSUGRA} is
 \eqn{CFback}{
  ds^2 = \e_0(z)^2 (-dt^2 + dx^2 + dy^2 + dz^2) \qquad\qquad
  \phi = \phi(z) \,,
 }
with all other fields set to $0$.  For example, if $V(\phi) = -6/L^2$, then
\begin{equation}
\label{E:AdS4}
	\e_0(z) = L/z  \qquad\qquad
	\phi(z) = 0
\end{equation}
provide a solution to the equations of motion. An solution \eqref{E:AdS4} is the Poincar\'e patch of AdS${}_4$, dual to a conformal field theory (CFT) in $2+1$ dimensions.  In the coordinate system used in \eqref{CFback} and \eqref{E:AdS4}, $z$ runs from $0$, which is the boundary of AdS${}_4$,  to infinity. Typically the CFT is a large $N$ gauge theory, and the large $N$ approximation justifies the classical treatment of the bulk dynamics.  The large $N$ gauge groups of the boundary theory have nothing to do with the bulk gauge group ${\bf g}$.  Rather, ${\bf g}$ is the algebra of continuous global symmetries of the boundary theory. We will denote the Noether currents associated with these symmetries by $J_m^a$, where $a$ is an index for the adjoint of ${\bf g}$ and $m$ runs over the three boundary directions.

There are more general solutions to the equations of motion following from \eqref{SSUGRA}. In what follows we will be interested in solutions of the form \eno{SSUGRA} where $\e_0$ and $\phi$ are defined for $z>0$, and where $\e_0$ is a monotonically decreasing function of $z$ which diverges at $z=0$.  Such solutions can frequently be related to vacua of conformal field theories deformed by an operator ${\cal O}_\phi$ dual to $\phi$.  When such a relation exists, the solutions are termed holographic renormalization group (RG) flows.  True to this name, these geometries break conformal invariance.  However, they manifestly preserve the Lorentz invariance of ${\bf R}^{2,1}$.  They also preserve the full gauge invariance under ${\bf g}$, because $\phi$ is (by assumption) a gauge singlet.

What we want particularly to note is that the solution \eno{CFback} (even when it's not AdS${}_4$) is unaltered upon the introduction of non-zero gauge field $A_\mu$, provided the field strength $F_{\mu\nu}$ vanishes.  More particularly, we are interested in configurations where $A_z=0$ and $A_m$ is constant for $m=0,1,2$.  In order to be a flat connection, $A_m$ must take values in a Cartan subalgebra ${\bf h}$ of ${\bf g}$:
 \eqn{AmDef}{
  A_m = A_m^a t^a \in {\bf h} \,.
 }
Through an appropriate choice of basis, we can insist that all the $t^a$ occurring in the sum \eno{AmDef} belong to ${\bf h}$.  We will refer to the flat gauge connection \eno{AmDef} as a Wilson line.  This would be more strictly appropriate if we compactified the $x$ and $y$ directions into a torus and formed the integrals $\int_\gamma A$ over closed curves $\gamma$ wrapping the torus.

The gauge fields \eno{AmDef} explicitly break some or all of the gauge and Lorentz invariance, while preserving translation invariance.  In the context of AdS/CFT, if the lagrangian of the field theory dual to the holographic RG flow \eno{CFback} is ${\cal L}_0$, then the lagrangian with the gauge fields \eno{AmDef} turned on is
 \eqn{Ldeformed}{
  {\cal L} = {\cal L}_0 + A_m^a J^{ma} \,.
 }
Two-point functions of operators which are singlets under ${\bf g}$, such as ${\cal O}_\phi$ and the stress tensor $T_{mn}$, respect the Lorentz invariance of the background \eno{CFback}, at least to leading order in $N$, because the free propagation of singlet fields, such as perturbations of $\phi$ and $g_{\mu\nu}$, do not respond to the Wilson lines \eno{AmDef}. On the other hand, correlation functions of the operators ${\cal O}_\Sigma$ and ${\cal O}_\Psi$ dual to $\Sigma$ and $\Psi$, and also of the Noether currents $J_m^a$, must be sensitive to the Wilson line.  The main goal of this whole section is to describe that dependence.

The key to making the calculation of the two-point functions of ${\cal O}_\Sigma$, ${\cal O}_\Psi$, and $J_m^a$ tractable is that the action simplifies greatly upon the conformal transformations 
 \eqn{ConfTrans}{
  \Sigma = {\sigma \over \e_0} \qquad
  \Psi = {\psi \over \e_0^{3/2}} \,.
 }
The parts of the action \eno{SSUGRA} involving $A_\mu$, $\Sigma$, and $\Psi$ can now be expressed as 
 \eqn{FlatAction}{
  S_{\rm flat} = \int_M d^4 x \, \left[ -{1 \over 2} \tr F_{\mu\nu}^2 -
     i \bar\psi \Gamma^{\underline{\mu}} D_\mu \psi - |D_\mu \sigma|^2 \right] -
    \int_{\partial M} d^3 x \, i \bar\psi \Gamma_- \psi \,,
 }
where we have now written $S_{\rm bdy}$ for the fermions explicitly in terms of the projection matrix $\Gamma_-$, and we have defined
 \eqn{GammaMinus}{
  \Gamma_\pm = {1 \mp \Gamma^{\underline{z}} \over 2} \,.
 }
$M$ is now the $z>0$ part of ${\bf R}^{3,1}$, and $\partial M$ denotes the surface $z=0$, which is just ${\bf R}^{2,1}$.  In \eno{GammaMinus}, and in the rest of this section, we use the flat metric $\eta_{\mu\nu} = \diag\{ -1,1,1,1 \}$ in all formulas: thus, for example, $\tr F_{\mu\nu}^2 = \eta^{\mu_1\mu_2} \eta^{\nu_1\nu_2} \tr F_{\mu_1\nu_1} F_{\mu_2\nu_2}$.  Because there is now no distinction between ``curved space'' indices $\mu$ and tangent space indices $\underline{\mu}$, we simply use $\mu$. Because the metric is flat, the spin connection vanishes.  So
 \eqn{DDefsAgain}{
  D_\mu\sigma &= (\partial_\mu - i \gYM A_\mu^a t^a) \sigma  \cr
  D_\mu\psi &= \left( \partial_\mu - i \gYM A_\mu^a t^a \right) \psi \,.
 }

\subsection{Scalar two-point functions}
\label{SCALARS}

We define
 \eqn{GreensDef}{
  G_\Sigma^R(x) &= -i\theta(t) \langle [ {\cal O}_\Sigma(t,\vec{x}),
    {\cal O}_\Sigma^\dagger(0,0) ] \rangle =
    \int {d^3 k \over (2\pi)^3} e^{ik \cdot x} G_\Sigma^R(k)  \cr
  G_\Sigma^F(x) &= -i \langle T {\cal O}_\Sigma(t,\vec{x})
    {\cal O}_\Sigma^\dagger(0,0) \rangle =
    \int {d^3 k \over (2\pi)^3} e^{ik \cdot x} G_\Sigma^F(k) \,.
 }
Here $T$ denotes time-ordering, $x = (t,\vec{x})$, $k = (\omega,\vec{k})$, and $k \cdot x = -\omega t + \vec{k} \cdot \vec{x}$.  We use $\langle A \rangle = \tr \rho A$ where $\rho$ is the density matrix of the state under consideration. When the temperature and chemical potentials are zero this reduces to the vacuum expectation value $\langle 0|A|0 \rangle$. The Green's functions \eno{GreensDef} can be extracted from solutions to the wave equation for $\sigma$, as we now explain in some detail.

The scalar equation of motion is
 \eqn{ScalarEOM}{
  D_\mu D^\mu \sigma = 0 \,.
 }
Because of the translation invariance in the $(t,\vec{x})$ directions, we can cast solutions in the form
 \eqn{SolForm}{
  \sigma(t,\vec{x},z) = e^{i k \cdot x} \hat\sigma(z) \,.
 }
A straightforward calculation shows that
 \eqn{DhatProperty}{
  D_\mu \sigma = e^{i k \cdot x} \hat{D}_\mu \hat\sigma
 }
where
 \eqn{DhatDef}{
  \hat{D}_m = i K_m \qquad \hat{D}_z = \partial_z
 }
and
 \eqn{KmDef}{
  K_m = k_m - \gYM A_m^a t^a \,.
 }
\eno{ScalarEOM} can be recast as
 \eqn{SEOMagain}{
  (\partial_z^2 - K_m K^m) \hat\sigma = 0 \,,
 }
so the solution is $\hat\sigma = e^{\pm Kz}v$, where $K = \sqrt{K^mK_m}$ and $v$ is an arbitrary vector in ${\bf r}_\Sigma$. 
Recall that $t^a$ is in the Cartan subalgebra ${\bf h}$ of ${\bf g}$ and we are working in a basis where all the $t^a$ are diagonal. Thus, the $K_m$ are diagonal, and one can define non-polynomial functions of the $K_m$. Indeed, let $v_\lambda$ be the weight space associated with weight vector $\lambda^a$, i.e., $t^a v_\lambda = \lambda^a v_\lambda$ for all $t^a \in {\bf h}$.  Then $K_m v_\lambda = (k_m - k_{m,\lambda}) v_\lambda$ where
 \eqn{kmlDef}{
  k_{m,\lambda} = \gYM A_m^a \lambda^a \,.
 }
When defining $K = \sqrt{K^m K_m}$ we need to specify which branch of the square root to pick.  To start, let the spatial part $\vec{k}$ of $k$ be large enough so that $k_m - k_{m,\lambda}$ is spacelike for all weights $\lambda$.  Then the action of $K$ on vectors in the $\lambda$ eigenspace is simply $Kv_\lambda = \sqrt{(k_m-k_{m,\lambda})(k^m-k^m_{\lambda})}v_\lambda \equiv \sqrt{(k-k_\lambda)^2}\, v_\lambda$, where we pick the plus sign on the square root. Thus, in this purely spacelike case, where $K^2$ is a positive definite matrix acting on ${\bf r}_\Sigma$, we choose $K$ also to be positive definite.  Then the solutions to \eno{SEOMagain} which decay far from the boundary take the form
 \eqn{hsChoose}{
  \hat\sigma = e^{-Kz} v \,.
 }

Still assuming that $K^2$ is positive definite, one may extract the Green's function in the following manner.  The on-shell action for the scalar is
 \eqn{Ionshell}{
  S_{\rm on-shell} &= -\int_M d^4 x \, |D_\mu \sigma|^2 =
     -\int_M d^4 x \, \left[ \partial_\mu (\sigma^\dagger D^\mu \sigma) -
      \sigma^\dagger D_\mu D^\mu \sigma \right]
    = \int_{\partial M} d^3 x \, \sigma^\dagger \partial_z \sigma \,,
 }
where in the second step we used the equations of motion, and in the third step we used Stokes' Theorem and remembered that $D^z = \partial_z$.  The basic premise of AdS/CFT is that the on-shell action is the generating functional for Green's functions of the boundary theory.  In the current instance, this means that
 \eqn{GreenIdent}{
  S_{\rm on-shell} = W_2[\sigma_{\rm bdy},\sigma_{\rm bdy}^\dagger]
    \equiv -\int_{\partial M} d^3 x_1 d^3 x_2 \,
     \sigma_{\rm bdy}^\dagger(x_1) G_\Sigma(x_1-x_2)
     \sigma_{\rm bdy}(x_2) \,,
 }
where $G_\Sigma$ is a two-point function of ${\cal O}_\Sigma$, and $\sigma_{\rm bdy}$ is simply $\sigma$ evaluated at the boundary $z=0$.   The alert reader will note that we have not specified whether \eno{Gresult} is a retarded or time-ordered Green's function. At the moment we do not need to because the Hermitian parts of $G_\Sigma^R$ and $G_\Sigma^F$ coincide, and we are doing the computation for $k_m$ such that $K^2$ is positive definite.  To recover the full Green's function we will use an appropriate  pole passing prescription to go to more general $K^2$.

Now consider the following linear combination of solutions to the equation of motion for $\sigma$:
 \eqn{sigmaSpecial}{
  \sigma = \xi_1 e^{i k \cdot x} e^{-Kz} v_1 +
    \xi_2 e^{i q \cdot x} e^{-Qz} v_2 \,,
 }
where $\xi_1$ and $\xi_2$ are complex numbers, and $Q$ is defined from the momentum $q_m$ as $K$ is from $k_m$.  The on-shell action is now sesquilinear in $\xi_1$ and $\xi_2$. Plugging the boundary limit of the specific form \eno{sigmaSpecial} into the last expression in \eno{GreenIdent}, one obtains immediately
 \eqn{ddW}{
  {\partial^2 W_2 \over \partial\xi_1 \partial\xi_2^*} =
    -(2\pi)^3 \delta^3(q-k) v_2^\dagger G_\Sigma(k) v_1 \,.
 }
On the other hand, plugging \eno{sigmaSpecial} into the last expression in \eno{Ionshell} leads to
 \eqn{ddS}{
  {\partial^2 S_{\rm on-shell} \over \partial\xi_1 \partial\xi_2^*} =
   \int_{\partial M} d^3 x \, v_2^\dagger e^{-iq \cdot x} (-K) e^{ik \cdot x} v_1 \,.
 }
Equating \eno{ddW} and \eno{ddS} leads to
 \eqn{Gresult}{
  G_\Sigma(k) = K \,.
 }
This is almost our final result.  The full retarded or Feynman Green's function can be obtained from \eqref{Gresult} by an appropriate pole passing prescription:
 \eqn{PolePassing}{\seqalign{\span\TR\qquad & \span\TT}{
  \omega \to \omega + i\epsilon & to obtain $G^R$  \cr
  \omega \to \omega(1 + i\epsilon) & to obtain $G^F$.
 }}
The subtleties of recovering the imaginary part of Green's functions from a supergravity action which is real have received significant attention \cite{Son:2002sd,Herzog:2002pc,Gubser:2008sz,Iqbal:2009fd}.  Here let us simply make the well-known observation that with the prescription \eno{PolePassing} for retarded Green's functions, the scalar wave-functions are infalling: on eigenspaces where $K_m$ acts as a timelike vector, $K$ is a negative imaginary number, meaning that $e^{-Kz} v = e^{ip z}v$ with $p > 0$.  In other words, the number flux of scalar quanta is away from the boundary, falling into the bulk.

Already from the simple expression \eno{Gresult} we can reach one of the main conclusions of this section: The spectral measure is supported in the timelike regions of light-cones in $k$-space, one through $k_{m,\lambda}$ for each weight $\lambda$ in the representation ${\bf r}_\Sigma$.  To see this explicitly, we can simply note that $G_\Sigma(k)$ acts as
 \eqn{Gsl}{
  G_{\Sigma,\lambda}(k) = \sqrt{(k-k_\lambda)^2}
 }
on the weight space associated with $\lambda$.  This part of the Green's function has an imaginary part precisely when $k-k_\lambda$ is timelike, i.e.~in the light-cone passing through $k_\lambda$.  The spectral measure is essentially this imaginary part.

We have not concerned ourselves with the overall normalization of the action.  If we had, we would have found an overall prefactor on $G_\Sigma$ which scales as $N^{3/2}$ in constructions based on $N$ coincident M2-branes.

Giving the scalar $\Sigma$ an explicit mass term $m$, instead of the conformal coupling $-{1\over 6}|\Sigma|^2$ implies that the dual operator $\mathcal{O}_{\Sigma}$ has dimension $\Delta$ where $\Delta(\Delta-3) = m^2 L^2$ and $L$ is the radius of curvature of AdS${}_4$. The calculation of the two-point functions proceeds almost identically to what we laid out above, except that it has to be done in AdS${}_4$ rather than flat space, because the mass term is not conformally invariant.  The radial parts of the wave-functions (without any conformal rescaling) come out proportional to $z^{3/2} {\bf K}_{\Delta-3/2}(Kz)$ instead of $z e^{-Kz}$ as in \eqref{hsChoose}.  The pole passing prescription \eno{PolePassing} will convert the modified Bessel function ${\bf K}_{\Delta-3/2}$ into an appropriate Hankel function when $K_m$ acts as a timelike vector.  When $\Delta \neq n+3/2$ with $n$ an integer, the result for the two-point function is $G_\Sigma(k) = c_\Delta K^{2\Delta-3}$, where $c_\Delta$ is a $k$-independent factor.  Evidently, \eno{Gresult} corresponds to the case $\Delta=2$.  When $\Delta = n+3/2$ the Green's function takes the form $\tilde{c}_\Delta K^{2\Delta-3} \log K$. The limit $2\Delta-3 \to n$ coincides with this form of the Green's function since $c_{\Delta}$ diverges in this limit.

\subsection{Spinor and vector two-point functions}
\label{SPINS}

Our results for the scalar Green's functions $G_\Sigma$ can be summarized in the following simple terms.  With no gauge field present in the bulk, conformal invariance dictates $G_\Sigma(k) = c_\Delta k^{2\Delta-3}$.  For $\Delta=2$, the same expression can be deduced more generally for conformally flat backgrounds.  The effects of the gauge field are simply to replace $k$ with $K$, where $K_m = k_m - \gYM A_m^a t^a$ as in \eno{KmDef}.

In light of these results, it is reasonable to guess that Green's functions for spinor operators and conserved currents in the presence of a flat connection follow from similar replacements $k_m \to K_m$.  More explicitly, the Green's functions we have in mind are defined as follows:
 \eqn{CorrelatorDefs}{
  G^R_\Psi(x) &\equiv -i \theta(t) \langle \{ {\cal O}_\Psi(t,\vec{x}),
    {\cal O}_\Psi^\dagger(0,0) \} \rangle =
    \int {d^3 k \over (2\pi)^3} e^{ik \cdot x} G^R_\Psi(k)  \cr
  G^{R,ab}_{mn}(x) &\equiv -i \theta(t) \langle [J^a_m(t,\vec{x}),J^b_n(0,0)] \rangle
    = \int {d^3 k \over (2\pi)^3} e^{ik \cdot x} G^{R,ab}_{mn}(k) \,.
 }
Time-ordered Green's functions would be defined similarly.  We usually omit the adjoint indices $a,b$.  ${\cal O}_\Psi$ may have any dimension $\Delta$, but because $J_m^a$ is conserved, its dimension must be $2$.  Our expectation is that
 \eqn{GreensGuess}{
  G_\Psi(k) = -f_\Delta {\gamma^m K_m \over K^{4-2\Delta}} \gamma^t \qquad\qquad
  G^{ab}_{mn}(k) = s_\infty \left( {K^2 \eta_{mn} - K_m K_n \over K} \right)^{ab} \,,
 }
where $f_\Delta$ and $s_\infty$ are constants that we do not propose to track at this stage, and the $i\epsilon$ prescription  \eno{PolePassing} is implied.  Our notation, essentially following \cite{Iqbal:2009fd}, is to represent the four-dimensional gamma matrices as
 \eqn{GammaExpress}{
  \Gamma^{{m}} = \begin{pmatrix} 0 & \gamma^m \\ \gamma^m & 0 \end{pmatrix}
    \qquad
  \Gamma^{{z}} = \begin{pmatrix} -1 & 0 \\ 0 & 1 \end{pmatrix} \,,
 }
where 
 \eqn{GammaChoice}{
  \gamma^t = i \sigma_2 \qquad \gamma^1 = \sigma_1 \qquad \gamma^2 = \sigma_3
 }
and $\sigma_a$ denotes the Paul matrices.  We 
decompose the Dirac spinor $\psi$ as
 \eqn{SpinorExpress}{
  \psi = \begin{pmatrix} \psi_+ \\ \psi_- \end{pmatrix} \,.
 }
The operator ${\cal O}_\Psi$ transforms as a two-component spinor, like $\psi_-$.  We sometimes consider the Dirac conjugate of a two component spinor: $\bar\psi_\pm \equiv \psi_\pm^\dagger \gamma^t$.  The Green's function  $G_\Psi(k)$ has the same spinor structure and group representation content as the bilinear $\psi_- \psi_-^\dagger$.

Just as in the case of scalar Green's functions, if $u_-$ is a two-component spinor which belongs to a weight space of ${\bf r}_\Psi$ with weight vector $\lambda$ then the action of $G_{\rm \Psi}$ on $u_-$ is of the form $G_\Psi(k) u_- = G_{\Psi,\lambda}(k) u_-$, where
 \eqn{GFspecial}{
  G_{\Psi,\lambda}(k) = -f_\Delta {\gamma^m (k_m-k_{m,\lambda}) \over
    (k-k_\lambda)^{4-2\Delta}} \gamma^t \,,
 }
and $k_\lambda$ is defined as in \eno{kmlDef}, only using weights of ${\bf r}_\Psi$ not ${\bf r}_\Sigma$.  Likewise, if $\epsilon^n$ is a polarization vector which also belongs to a root space of ${\bf g}$ with root vector $\alpha$, then $G_{mn}(k) \epsilon^n = G_{mn,\alpha}(k) \epsilon^n$ where
 \eqn{GJspecial}{
  G_{mn,\alpha}(k) = s_\infty {(k-k_\alpha)^2 \eta_{mn} -
   (k_m-k_{m,\alpha})(k_n-k_{n,\alpha}) \over \sqrt{(k-k_\alpha)^2}} \,,
 }
and $k_{m,\alpha} = \gYM A^a_m \alpha^a$.  Evidently, the spectral measures of $G_\Psi$ and $G_{mn}$ are supported inside light cones, just as for $G_\Sigma$, but each light cone has its apex at the the momentum $k_\lambda$ or $k_\alpha$ associated with a weight vector $\lambda$ or a root vector $\alpha$, as appropriate.

Let's sketch a derivation of the expression in \eno{GreensGuess} for $G_\Psi$ in the case $\Delta = 3/2$, corresponding to a massless fermion, again focusing first on the situation where $K^2$ is positive definite.  This is the case where we can do all the calculations in flat space, where the equation of motion is
 \eqn{Psieom}{
  \Gamma^\mu D_\mu \psi = 0 \,.
 }
This equation implies $D_\mu D^\mu \psi = 0$, which is identical to \eno{ScalarEOM}.  So the allowed solutions are of the same form:
 \eqn{PsiSolns}{
  \psi = e^{ik \cdot x} e^{-Kz} u \,.
 }
Here $u$ is a four-component spinor transforming in the representation ${\bf r}_\Psi$.  Plugging \eno{PsiSolns} into \eno{Psieom} provides a constraint on $u$:
 \eqn{DiracConstraint}{
  \left( i\Gamma^m K_m - \Gamma^z K \right) u = 0 \,.
 }
Decomposing $u$ as in \eno{GammaExpress}, one can rewrite \eno{DiracConstraint} as
 \eqn{DCA}{
  u_- = i {\gamma^m K_m \over K} u_+ \,.
 }
 
To compute the fermionic Green's function we plug a linear combination of solutions to the Dirac equation,
 \eqn{PsiSpecial}{
  \psi = \xi_1 e^{ik\cdot x} e^{-K z} u_1 +
    \xi_2 e^{iq\cdot x} e^{-Q z} u_2 \,,
 }
into the on-shell action, which is simply the boundary term in \eno{FlatAction}:
 \eqn{SPsiOnshell}{
  S_{\rm on-shell} = -\int_{\partial M} d^3 x \, i \bar\psi \Gamma_- \psi =
    -i \xi_2^* \xi_1 (2\pi)^3 \delta^3(q-k) \bar{u}_{2+} u_{1-} + \ldots \,.
 }
In \eqref{SPsiOnshell} we omitted terms involving different combinations of the $\xi_i$ and their conjugates.  Using \eno{DCA} we find
 \eqn{ddSPsi}{
  {\partial^2 S_{\rm on-shell} \over \partial\xi_1 \partial\xi_2^*} =
    (2\pi)^3 \delta^3(q-k) \bar{u}_{2+} {\gamma^m K_m \over K} u_{1+} \,.
 }
On the other hand, essentially by definition,
 \eqn{ddWPsi}{
  {\partial^2 W_2 \over \partial\xi_1 \partial\xi_2^*} =
    -(2\pi)^3 \delta^3(q-k) \bar{u}_{2+} G_\Psi(k) \gamma^t u_{1+} \,.
 }
Comparing \eno{ddSPsi} and \eno{ddWPsi} leads immediately to
 \eqn{GotGPsi}{
  G_\Psi(k) = -{\gamma^m K_m \over K} \gamma^t \,.
 }
Continuing to momenta where $K^2<0$ can be done using \eno{PolePassing}.

For the gauge field, the story is essentially the same.  We start by perturbing the gauge field:
 \eqn{Aperturbed}{
  A = A_{\rm flat} + a \,.
 }
The field strength is $F = {i \over g} (d - i gA)^2 = f + O(a^2)$, where
 \eqn{SmallF}{
  f_{\mu\nu} = D_\mu a_\nu - D_\nu a_\mu \,,
 }
where by convention $D = d - i g A_{\rm flat}$.  The linearized equation of motion is
 \eqn{aeom}{
  D^\mu f_{\mu\nu} = 0 \,.
 }
It is most straightforward to proceed in a gauge where $a_z=0$.  Then one can show directly from \eno{aeom}, assuming as usual that $K^2$ is positive definite, that $a_m = e^{ik_m x^m} e^{-Kz} \epsilon_m$ if $K^m \epsilon_m = 0$, and that $a_m = e^{ik_m x^m} \epsilon_m$ if ${K_n K^m \over K^2} \epsilon_m = \epsilon_n$. Here $\epsilon_n$ is a constant polarization vector.  The transverse modes, where $K^m \epsilon_m = 0$, are physical and the others are pure gauge. If we define
 \eqn{PmnDef}{
  P_n{}^m = \delta_n{}^m - {K_n K^m \over K^2} \,,
 }
then we may compactly write the allowed solutions as
 \eqn{aSolns}{
  a = e^{ik \cdot x} e^{-K P z} \epsilon \,,
 }
where Lorentz indices for the boundary directions are now implied.  Plugging \eno{aSolns} into the quadratic on-shell action
 \eqn{Squad}{
  S_{\rm on-shell} = \int d^3 x \, \tr a_m f^{zm} \,,
 }
one finds more or less immediately that
 \eqn{GotGK}{
  G_{mn}(k) = K P_{mn} \,,
 }
which indeed has the form indicated in \eno{GreensGuess}.

We will not delve deeply into conductivity calculations in this paper, but it is worth noting that \eno{GotGK} contains a geometrical explanation of the hard-gap phenomenon remarked on in \cite{Basu:2009vv}.  To understand this connection, first recall that the gauge potential in the infrared copy of AdS${}_4$ is (in the notation of \cite{Basu:2009vv}) $A = B_0 \, \tau^1 \, dx$.  So the Cartan subalgebra ${\bf h}$ is the one generated by $\tau^1$.  The conductivity studied in \cite{Basu:2009vv} is the one related to a perturbation $\delta A = e^{-i\omega t} a(z) \, \tau^3 \, dy$.  Now, $G_{mn}^R(k)$ acts in the adjoint representation, so it can be denoted $G_{mn}^{R\;ab}(k)$.  The conductivity of interest is 
 \eqn{BasuConductivity}{
  \sigma_{yy,\tau^3}(\omega) = {i \over \omega} v_a v_b G_{yy}^{R\;ab}(\omega,0,0) \,,
 }
where $v_a$ is the unit vector corresponding to the generator $\tau^3$.  Now we need to find a convenient basis in which to work out the right-hand-side of \eno{BasuConductivity}.  The obvious basis for the adjoint representation, for our purposes, is the one in which $\tau^1$ acts diagonally.  The eigenvalues of its adjoint action are $\alpha = \pm 1$ and $0$: these are the roots $\alpha$ of the adjoint representation of $SU(2)$, and they correspond to the c-number Green's functions $G_{mn,\alpha}^R$ defined in \eno{GJspecial}.  In the basis for the adjoint representation where $(1,0,0)$ corresponds to $\alpha=1$, $(0,1,0)$ corresponds to $\alpha=-1$, and $(0,0,1)$ corresponds to $\alpha=0$, the unit vector corresponding to the generator $\tau^3$ is $v = (1,1,0)/\sqrt{2}$.  Plugging this expression for $v$ into \eno{BasuConductivity}, we have
 \eqn{FoundBC}{
  \sigma_{yy,\tau^3}(\omega) = {i \over 2\omega} \left[ G_{yy,+}^R(\omega,0,0) + 
    G_{yy,-}^R(\omega,0,0) \right] \,.
 }
Now we can explain the geometric origin of the gap. From the explicit expression \eno{GJspecial} we see that $G_{yy,+}^R$ is real except inside the light cone whose apex is at $k_{m,+} = \gYM B_0 \delta^x_m$.  The distance along the $\omega$ axis one must go in order to cross into this light cone is $\gYM B_0$.  Likewise, $G_{yy,-}^R$ is real except inside a lightcone which one crosses into along the $\omega$ axis at the same value of $\omega$.  $G_{yy,0}^R$ has no gap, but it is not involved in $\sigma_{yy,\tau^3}(\omega)$.  So the gap is $\Delta = \gYM B_0$.

What we are neglecting is that there is asymmetric warping of time and space in the domain wall geometry of \cite{Basu:2009vv}.  The $t$ coordinate picks up a warp factor $e^{-\chi_0/2}$ in the infrared AdS${}_4$ geometry relative to the ultraviolet, while the $x$ coordinate picks up a factor $c_0$.  Correspondingly, the wave-vector $k_{m_+}$ at the apex of the light cone gets scaled by $1/c_0$ because it points in the $x$ direction, and frequency gets scaled by $e^{\chi_0/2}$.  Altogether, introducing these rescalings into the expression $\Delta = \gYM B_0$ for the gap leads to
 \eqn{FoundBG}{
  \tilde\Delta = {e^{-\chi_0/2} \gYM B_0 \over c_0} \,.
 }
$\tilde\Delta$ should match the gap found in \cite{Basu:2009vv}, and it does.  (Recall that their $q$ is our $\gYM$.)  What we learn in addition is that if conductivity could be measured at finite wave-number, the gap would decrease as one goes toward the apex of one of the lightcones, and can be reduced to zero if the wave-number is chosen to lie at the apex.  Similar analysis seems possible for $\sigma_{xx,\tau^3}$, but we have not fully considered the consequences of mixing with the graviton and with timelike components of the gauge potential in this case.

It may seem striking that we are able to evaluate the gap by considering only the infrared limit of the geometry.  This actually makes sense because there is a continuum contribution to the spectral measure of $G_{mn}^{R\;ab}$ only when the corresponding gauge-boson wave-function has some infalling component in the infrared geometry; otherwise the Green's function is purely real, with no dissipative part, except for $\delta$-function localized contributions to the spectral measure like the one that signals infinite DC conductivity.  The same reasoning, essentially, is behind the evaluation of the gap in \cite{Basu:2009vv}: the Schrodinger potential plateaus in the infrared at an energy that determines the gap.

\subsection{Summary and an example}
\label{SUMMARY}

The upshot of the calculations in this section is that the Green's functions of operators dual to the conformally coupled scalar $\Sigma$, the massless Dirac fermion $\Psi$, and the non-abelian gauge fields $A_\mu$ in \eqref{SSUGRA}, are simple functions of modified momenta
 \eqn{KmDefAgain}{
  K_m = k_m - \gYM A_m^a t^a \,.
 }
In particular, they all involve fractional powers of $K^2 = \eta^{mn} K_m K_n$.  $K^2$ is always hermitian, but positive definite only for $k$ outside all the shifted light-cones $(k_m-k_{m,\lambda})(k^m-k^{m}_{\lambda})=0$. As explained earlier, $k_{m,\lambda}$ is given by
 \eqn{kmlDefAgain}{
  k_{m,\lambda} = \gYM A_m^a \lambda^a \,,
 }
where $\lambda$ is a weight vector for the representation in which the operator whose Green's function we are considering transforms.  The spectral measure of these Green's functions comes entirely from the region where $K^2$ fails to be positive definite, i.e. the region where $k-k_\lambda$ is timelike for some weight $\lambda$.

It is interesting that the Green's functions have a pure power law form not only for an AdS${}_4$ bulk, but for any conformally flat bulk geometry which is asymptotically AdS${}_4$.  This is a consequence of the invariance of the relevant part of the classical supergravity action under conformal transformations.  Thus, power-law correlators, like those one sees in conformal theories, arise at leading order in a large $N$ limit for special operators in backgrounds that are dual to non-conformal theories.  At subleading orders in $N$, loop corrections in the bulk enter in, and one cannot expect the exact power-law behavior of the boundary theory correlators to persist unless there is conformal symmetry.  All the backgrounds treatable by the methods of this section have exact boost invariance, $SO(2,1)$, as well as translational invariance.  These symmetries would be preserved at all loop orders in the bulk, meaning all orders in $N$ in the boundary theory.

As an example, consider ${\bf g} = so(4) = su(2)_A \oplus su(2)_B$, where the generators for the two $su(2)$ factors are labeled $\tau_A^a$ and $\tau_B^a$.  The standard Cartan subalgebra is generated by $\tau_A^3$ and $\tau_B^3$.  Let the fermions $\Psi$ transform in the $4 = 2_A \times 2_B$ of $so(4)$.  Consider now the flat connection
 \eqn{SOfourFlat}{
  A = W \, (dx \, \tau_A^1 + dy \, \tau_B^2) \,,
 }
where $W$ is a constant.  Assuming that for $X=A,B$ the eigenvalues of each $\tau_X^a$ acting on the doublet of $su(2)_X$ are $\pm 1/2$, the images \eno{kmlDefAgain} of the weight vectors of the vector representation of $so(4)$ are
 \eqn{klChoices}{
  k_{s_1s_2} = {\gYM W \over 2} (s_1 \hat{k}_1 + s_2 \hat{k}_2)
 }
where $s_i=\pm 1$ and $\hat{k}_i$ are unit vectors in momentum space.
Thus the apexes of the Dirac cones pass through points at the corner of a square whose sides are aligned with the $k_x$ and $k_y$ axes.

\begin{figure}
\begin{center}
\includegraphics[height = 8 cm]{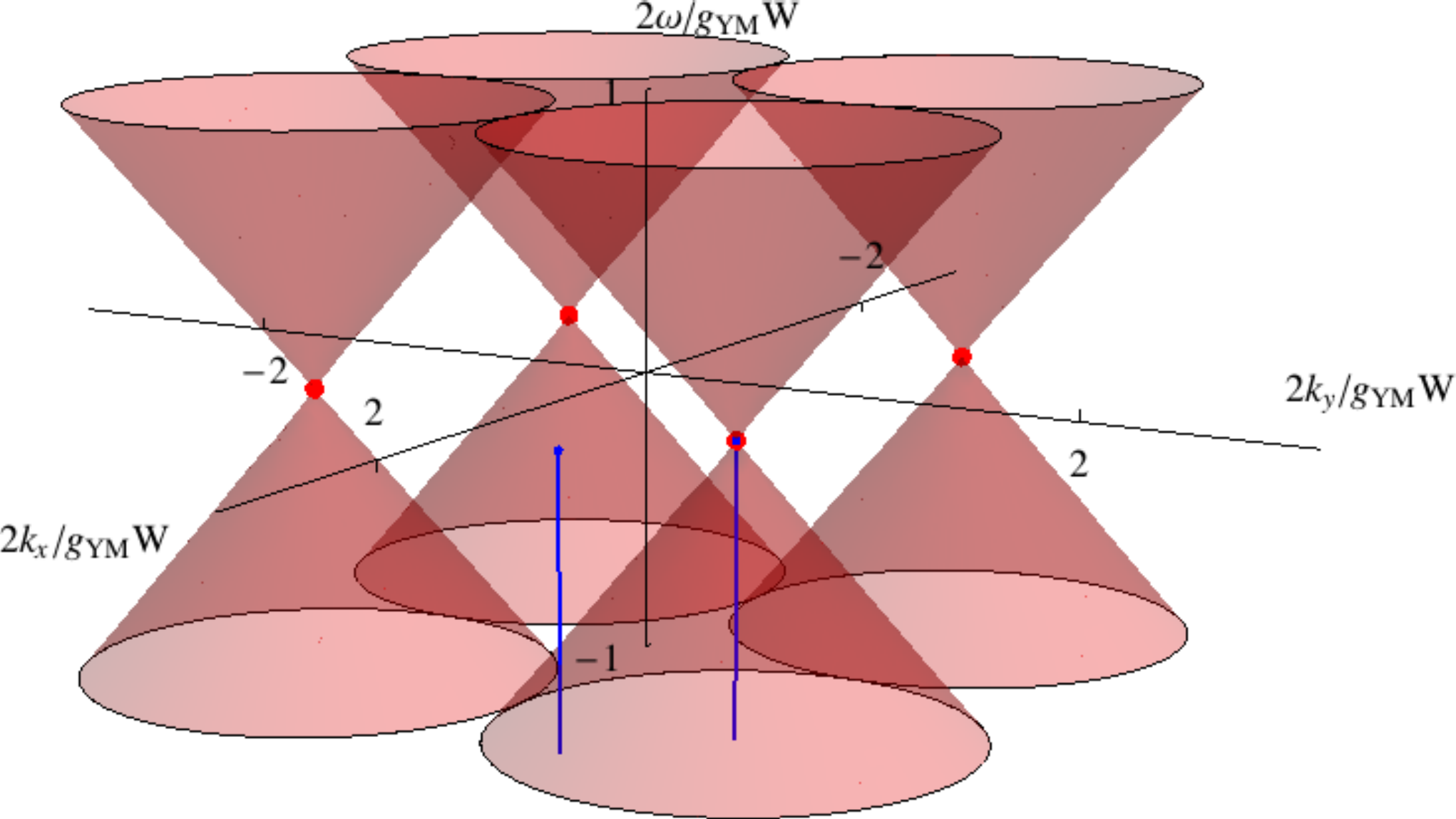} \\
\includegraphics[width = 6.5 in]{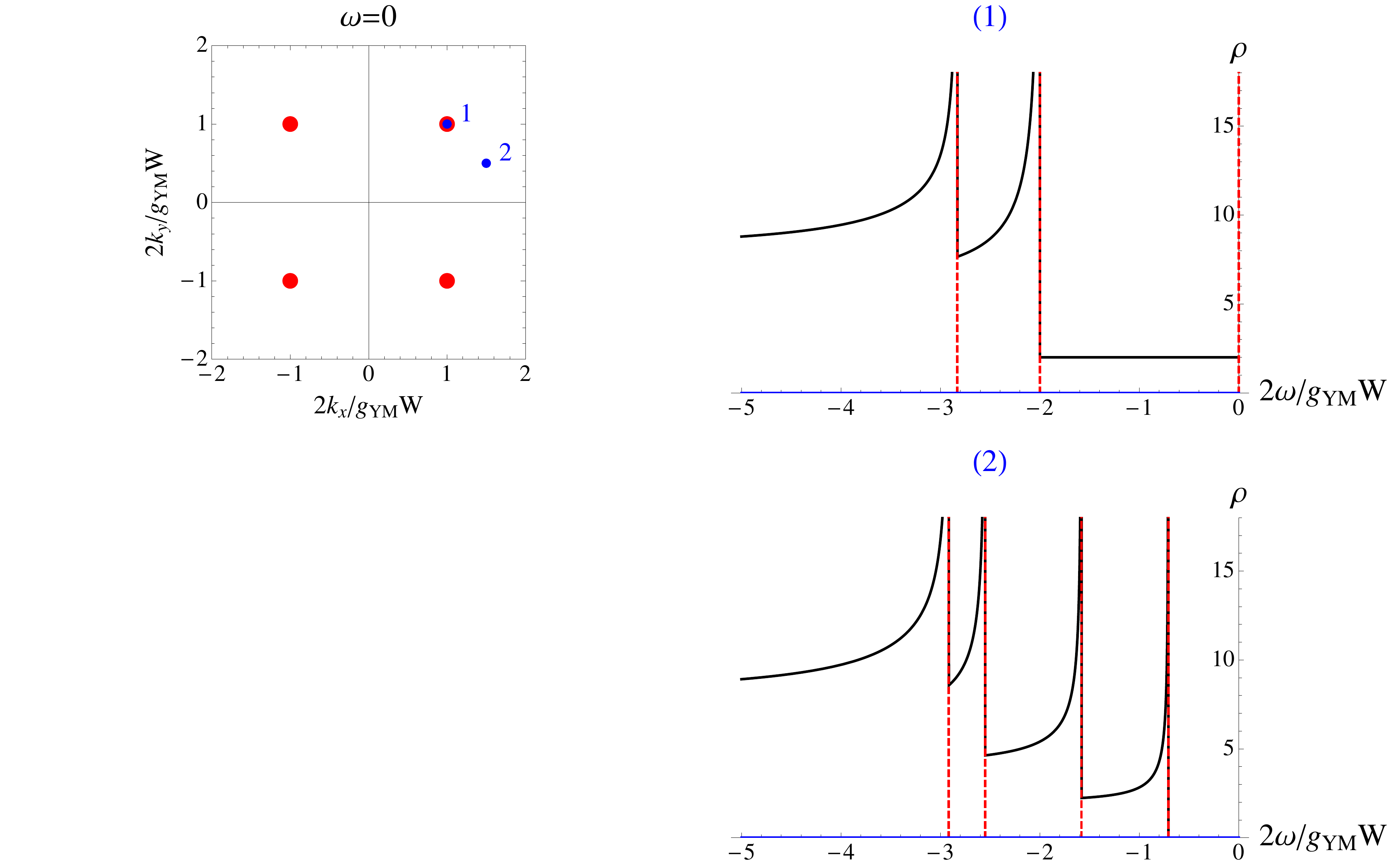}
\end{center}
\caption{Features of the spectral function for the operator dual to a massless Dirac fermion in the ${\bf 4}$ of $so(4)$, in a conformally flat background with the flat $so(4)$ connection \eno{SOfourFlat}.  This is distinct from the spectral function in the $p$-wave superconductor to be discussed in the next section. Top: The Dirac cones inside which the spectral measure is supported.  Bottom: Two representative plots of $\rho(\omega,\vec{k})$ as a function of $\omega$ for fixed values of $\vec{k}$.}\label{IRPLOTS2}
\end{figure}

The spectral measure of the retarded version of the Green's function \eno{GotGPsi} is 
 \eqn{rhoDef}{
  \rho(\omega,\vec{k}) \equiv -\Im\tr G^R_\Psi(k) = 2 |\omega| \sum_{s_i = \pm 1}
    {\theta(\omega^2-|\vec{k}-\vec{k}_{s_1s_2}|^2) \over
      \sqrt{\omega^2-|\vec{k}-\vec{k}_{s_1s_2}|^2}} \,,
 }
where we have used \eno{GotGPsi} and \eno{klChoices}.  The most conspicuous feature of \eno{rhoDef} is that $\rho(\omega,\vec{k})$ is supported in the union of four Dirac cones, whose apexes are the vectors $\vec{k}_{s_1s_2}$.  Plots of the Dirac cone and of $\rho(\omega,\vec{k})$ as a function of $\omega$ for fixed $\vec{k}$ are shown in figure~\ref{IRPLOTS2}.  By adding a mass to the fermion, we can adjust the power in the denominator of \eno{rhoDef} from a square root to $2-\Delta$, where $\Delta$ is the dimension of the dual fermionic operator in the field theory: see \eno{GFspecial}.

\section{Properties of the spectral function}
\label{SPECTRAL}

We are interested in studying the Fourier space retarded Green's function $G^R(k^m)$ for  fermions in the doublet of $SU(2)$ for the zero temperature $p$-wave superconductors discussed in section~\ref{PWAVE}; Our conventions for $G^R$ can be found in \eqref{CorrelatorDefs}. Note that $G^R$ has both spinor and gauge indices and a total of 16  \emph{a priori} independent components.  Here we will be interested in the spectral measure, $\rho$, defined by $\rho=-\Im\tr G^R$, where the trace is over both the gauge indices and the spinor indices.  This quantity is gauge invariant, and it is interesting to consider because it gives a measure of the number of eigenstates  of the theory that couple to the fermionic operator, with long lived excitations manifesting themselves as $\delta$-functions peaks.  In what follows we explain in some detail how to compute the spectral function (section \ref{METHOD}), how it relates to normal modes of bulk fermions (section \ref{NORMAL}) and how positivity of $\rho$ is guaranteed from a bulk point of view (section \ref{POSITIVITY}). The reader interested in the final numerical results for $\rho$ may skip directly to section \ref{FERMIONS}.

\subsection{Green's functions from a gravity dual}
\label{METHOD}

In section \ref{CONFORMAL} we computed the retarded Green's function for a configuration with a flat connection. In this section we consider the fermion Green's function for the zero temperature $p$-wave superconductor discussed in section \ref{PWAVE} where the connection is flat only in the IR and UV.  The action for the background is given by \eqref{E:PwaveAction} and we use the notation in (\ref{AforSU2}-\ref{gforSU2}) to describe the metric and $SU(2)$ gauge fields. We introduce the spin-$1/2$ field $\Psi$ which transforms in the doublet of $SU(2)$, whose action is given by
\begin{equation}
\label{SFermionTwice}
S_{\rm fermion} = -i \int_M d^4x \sqrt{-g}\, \bar{\Psi} \Gamma^{\mu}D_{\mu}\Psi-i\int_{\partial M} d^3x\,\sqrt{-g g^{rr}}\, \bar{\Psi}\Gamma_-\Psi\ \,,
\end{equation}
where $D_\mu$ is given in \eqref{DDef}.  Our conventions for the $\Gamma$ matrices are largely as in section~\ref{SPINS}, but because we now work in curved spacetime, we must be careful to distinguish between curved and flat indices.  Because the matrices $\gamma^m$ are needed to describe the physics of the boundary theory, which is defined on a flat background, we persist in defining them as in \eno{GammaChoice}.  In place of \eno{GammaExpress} we employ
 \eqn{GammaNewExpress}{
  \Gamma^{\underline{m}} = \begin{pmatrix} 0 & \gamma^m \\ \gamma^m & 0 \end{pmatrix}
    \qquad
  \Gamma^{\underline{r}} = -\Gamma^{\underline{z}} = \begin{pmatrix} 1 & 0 \\ 0 & -1 \end{pmatrix} \,.
 }
The second term in \eqref{SFermionTwice} is a boundary term that does not affect the equations of motion but gives the only nonzero contribution to the on-shell action.  We treat $\Psi$ in the probe approximation and do not allow it to back-react on the geometry. The equation of motion for $\Psi$ is $D_\mu \Psi=0$. To write it explicitly, it is convenient to exploit translation invariance in the $x^m$ directions and write $\Psi$ in the form
\eqn{psiDefGen}{
\Psi(x^m, r) = (-g g^{rr})^{-{1\over 4}} e^{-i \omega t + i k_x x + i k_y y} \psi(r) \,.
}
Note that \eqref{psiDefGen} reduces to \eqref{ConfTrans} for a conformally flat metric. As in section~\ref{CONFORMAL}, it is useful to split $\psi$ into two chiral spinors
\eqn{psipm}{
  \psi = \begin{pmatrix} \psi_+ \\ \psi_- \end{pmatrix} \,.
}
The equations of motion for $\psi_{\pm}$ take the form:
\eqn{psiEOM}{
\psip' + { i\over \sqrt{\gamma}r^2 } \left[
{e^{\chi\over 2}\over \sqrt{\gamma}} \left(-\omega - \gYM \PHI\, \tau^3\right)\gammat +
{1\over c} \left(k_x - \gYM W\, \tau^1\right)\gamma^{ x}
+ k_y \gamma^{ y}
\right] \psim
&= 0\cr
-\psim' + {i\over \sqrt{\gamma}r^2 } \left[
{e^{\chi\over 2}\over \sqrt{\gamma}} \left(-\omega - \gYM \PHI\,\tau^3\right)\gammat +
{1\over c} \left(k_x - \gYM W\tau^1\right)\gamma^{ x}
+ k_y \gamma^{ y}
\right] \psip
 &= 0\,,
}
where the prime denotes a derivative with respect to $r$ and $\gamma^{{\mu}}$ are $2+1$ dimensional boundary theory $\gamma$ matrices related to the bulk $\Gamma$ matrices through \eqref{GammaExpress}. Note that $\psi_\pm$ have both a boundary spinor index and an $SU(2)$ doublet index. We suppressed both types of indices in \eqref{psiEOM} but it should be clear how they are contracted. For instance $\left(\gamma^t \tau^1\psi_+\right)^{(i)\,(b)}=\gamma^{ t\,(ij)} \tau^{1\,(ab)} \psi^{(j)\,(b)}$, where $(i),(j),\ldots=1,2$ are spinor indices and $(a),(b),\ldots=1,2$ are $SU(2)$ doublet indices.

The asymptotic solution to \eqref{psiEOM} in the infrared is
\eqn{psiIR}{
\psi^{\rm IR} &= V_1 e^{-\kappa_+/r}+V_2 e^{-\kappa_-/r} + V_3 e^{\kappa_+/r}+V_4 e^{\kappa_-/r}
  \cr
\kappa_{\pm} &\equiv \left(-e^{\chi_{\rm IR}} \omega^2 + \left(2k_x \pm \gYM W_{\rm IR} \over 2c_{\rm IR}\right)^2+k_y^2\right)^{1\over 2}\,,
}
Here $c \IRarrow c_{\rm IR}$ and similarly $W \IRarrow W_{\rm IR}$ and $\chi \IRarrow \chi_{\rm IR}$. The spinor integration constants $V_1$ and $V_2$ satisfy $\tau^1 V_i=\pm {1\over 2} V_i$ with the minus sign for $i=1,3$ and the plus sign for $i=2,4$. They must also satisfy a linear constraint  that follows from \eqref{DiracConstraint}. The equations $\kappa_\pm=0$ define the two IR Dirac cones that we expect for $SU(2)$ doublet fermions. As discussed in section \ref{CONFORMAL} if we are outside one of the Dirac cones, i.e, if  for instance $\kappa_+^2>0$, then the corresponding term in \eqref{psiIR} should be regular provided we pick the positive sign for the square root in the definition of $\kappa_+$. If we are inside one of the Dirac cones, the correct boundary conditions for computing the retarded Green's function are obtained by making the substitution $\omega\to \omega + i \epsilon$ and demanding regularity. In practice, this prescription amounts to keeping $\omega$ real and picking the sign of the square root such that the imaginary part has the same sign as $\omega$. This is equivalent to requiring that the solutions are infalling in the IR. Thus, we set
\begin{equation}
\label{IRregularity}
	V_3 = V_4 = 0\,.
\end{equation}

The asymptotic behavior of the fermions in the UV takes the form
\eqn{psiUV}{
\psi^{\rm UV} = Q_1 e^{-\lambda_+/r}+Q_2 e^{\lambda_+/r}+Q_3 e^{-\lambda_-/r}+Q_4 e^{\lambda_-/r}
\cr\qquad \lambda_{\pm} \equiv \left(-e^{\chi_{\rm UV}} \left(\omega\pm {1\over 2}\gYM \PHI_{\rm UV}\right)^2 + {k_x^2\over c_{\rm UV}^2}+k_y^2\right)^{1\over 2}\,,
}
where $c \UVarrow c_{\rm UV}$ and a similar definition for $\chi_{\rm UV}$ and $\Phi_{\rm UV}$. The $Q_i$ are constant spinors such that $\tau^3 Q_i=\pm{1 \over 2}Q_i$ with the plus sign for $i=3,4$ and the minus for $i=1,2$. Like $V_1$ and $V_2$, each $Q_i$ satisfies a linear constraint that follows from  \eqref{DiracConstraint}.
As opposed to the IR asymptotics \eqref{psiIR}, in the UV there is no restriction on the choice of the sign of the square root in the definition of $\lambda_{\pm}$.
The equations $\lambda_\pm=0$ define two UV Dirac cones, different from the IR Dirac cones $\kappa_\pm=0$. The IR Dirac cones are shifted in the $k_x$ direction because in the IR only $A_x$ is nonzero. The UV Dirac cones are shifted in the $\omega$ direction since in the UV only $A_t$ is nonzero. Furthermore, since $c$ and $\chi$ go to different constants in the UV and IR, the UV and IR Dirac cones are stretched by different factors in the $\omega$ and $k_x$ directions: that is, they are characterized by different, anisotropic speeds of light.

The Green's function for the operator dual to $\Psi$ can be found from the UV behavior of the fermions following the prescription given in \cite{Henningson:1998cd,Mueck:1998iz} and more recently in \cite{Iqbal:2009fd}. For the purpose of computing the Green's function we rewrite the UV asymptotics in the form
\eqn{psiUVABCD}{
\psi_+^{\rm UV} &= A + B {1\over r} + \mathcal{O}\left({1\over r^2}\right) \cr
\psi_-^{\rm UV} &= D + C {1\over r} + \mathcal{O}\left({1\over r^2}\right)  \,,
}
where $A$, $B$, $C$ and $D$ are arbitrary chiral spinors with an $SU(2)$ doublet index which are linearly related to the $Q_i$. The linear constraints that follow from \eqref{DiracConstraint} can be used to solve for $B$ and $C$  in terms of $D$ and $A$. Since the equations of motion are linear, requiring that the solutions are regular in the IR, \eqref{IRregularity}, imposes a linear relation between $A$ and $D$,
\eqn{Mdef}{
D^{(i) (a)} = {\mathcal M}^{(ij) (ab)} A^{(j) (b)}\,.
}
As explained in section \ref{CONFORMAL}, to obtain $G^R$ we take the functional derivative of the on-shell action. The result is
\eqn{GotGRM}{
G^{R}= - i \mathcal{M} \gammat \,,
}
where we suppressed the spinor and $SU(2)$ indices. The derivation is very similar to the one for uncharged fermions, and we refer the reader to \cite{Iqbal:2009fd} for details.

\subsection{Normal modes}
\label{NORMAL}
We are particularly interested in the regions in which $\rho \neq 0$. Consider the IR asymptotics for the fermions, \eqref{psiIR}. If $\kappa_+^2 > 0$ or $\kappa_-^2 >0$ (i.e., we are inside the IR light-cone) then $\psi_+$ can be chosen to be everywhere real and consequently, $\rho=0$ except perhaps for poles of $\Re \rm{tr} (G_R)$ which, together with the $i\epsilon$ prescription corresponds to $\rho$ having $\delta$-function support. 
In what follows, we argue that a divergence of $G_{\rm R}$ corresponds to a solution to \eqref{psiEOM} which is a normal mode.

We define a normal mode to be a solution of \eqref{psiEOM} which is regular in the IR and its near boundary expansion takes the form \eqref{psiUVABCD} with $A=0$. This is a natural definition of a normal mode since $A$ is the leading term in the UV asymptotic expansion of $\psi$. We can not require $D$ to vanish as well since we are not free to choose $D$ once we fix $A$. Normal modes will only occur for special values of $k^m$ and $V_i$ (i.e., the values of $k^m$ for which there are normal modes will form a codimension one surface). 
The coefficients $A$ and $D$ in the UV expansion of $\psi$ and $V_1$ and $V_2$ in the IR expansion of $\psi$ are both integration constants of the same solution to a linear differential equation. Hence,
\eqn{UVDef}{
A= {\mathcal U}\,V \qquad D = {\mathcal V} \,V\,,
}
where $V$ should be understood as a four component object that contains the four independent integration constants in the $V_i$. Using \eqref{Mdef} the four by four matrices $\mathcal{U}$ and $\mathcal{V}$ are related to $\mathcal{M}$ through
\eqn{GotMUV}{
\mathcal{M} = \mathcal{V}\,\mathcal{U}^{-1}\,,
}
where some care must be taken in interpreting the index structure. 
Since $V$, $A$ and $D$ are finite the entries of $\mathcal{U}$ and $\mathcal{V}$ can not diverge. From here it follows that if $G^{R}$ has diverging entries then $\det\,\mathcal{U}=0$; the only way that $\mathcal{M}$ (and hence $G^{R}$) can diverge is if $\mathcal{U}$ is not invertible, i.e., the solution is a normal mode.

By the same reasons as argued in \cite{Gubser:2009dt}, we expect normal modes only outside  the IR Dirac cones. Let us briefly repeat this reasoning.  If $\kappa_\pm^2>0$ then, as discussed earlier, $\psi_+$ can be chosen to be everywhere real and the condition $\det\,\mathcal{U}=0$ is a real equation for which we can expect to find solutions for appropriate $\omega,\,k_x$ and $k_y$. On the other hand, if $\kappa_\pm^2<0$, $\psi_+$ cannot be chosen real and $\det\,\mathcal{U}=0$ will be a complex equation whose solutions are expected to involve complex $\omega$. Such a solution would correspond to a quasinormal mode rather than a normal mode.

There is also a somewhat weaker argument for expecting the normal modes to be inside at least one of the UV Dirac cones, ($\lambda_+^2<0$ or $\lambda_-^2<0 $). 
According to \eqref{psiUV}, if $\lambda_+^2>0$ and $\lambda_-^2>0$ then the generic UV solution will grow exponentially 
and matching such a solution to the IR asymptotics 
seems improbable.
If at least one of $\lambda_+^2$ and $\lambda_-^2$ is negative, then there is at least a two dimensional subspace of oscillatory solutions and there is no such obstruction. 
Our numerics indicate that the outermost UV light cone is indeed where the surface of normal modes ends.
We conclude that the normal modes exist in a ``preferred region'' 
\eqn{GotRegion}{
\kappa_+^2 >0\ \textrm{and}\ \kappa_-^2 >0\ \textrm{and}\ \Big(\lambda_+^2<0\ \textrm{or}\ \lambda_-^2<0\Big)\,.
}
This preferred region is bounded provided the UV Dirac cones have a narrower opening angle in both the $k_x$ and $k_y$ directions than the IR Dirac cones.  This is certainly true of the $SU(2)$ backgrounds we have constructed.   It may be possible to demonstrate in general that the region \eno{GotRegion} is compact starting from an appropriate positive energy condition.

\subsection{Positivity of the spectral measure}
\label{POSITIVITY}

Unitarity requires that the spectral measure  $\rho=-\Im\tr G_R$ is nonnegative for all real $k^m$ in any field theory, as can be shown using a spectral decomposition. Instead of explaining the spectral decomposition argument (which is standard) we will show in this section how the positivity of the spectral measure follows from a computation in the gravity dual.

Consider the current
\eqn{GotJmu}{
J^\mu \equiv
-\bar\Psi \Gamma^\mu \Psi \,.
}
which is conserved in the sense that $\nabla_\mu J^\mu=0$ provided the equations of motion that follow from \eqref{SFermionTwice} are obeyed.
Not surprisingly, this conserved current is associated to the gauge invariance of the action \eqref{SFermionTwice}. Here $\nabla$ is a covariant derivative.
From \eqref{psiDefGen}, we can write $\nabla_\mu J^\mu = 0$ as $Q'(r)=0$, where
\begin{equation}
\label{GotQ}
	Q(r) \equiv \sqrt{-g} J^r = -\bar\psi \Gamma^{\underline r} \psi 
		= -\bar\psi_+ \psi_- + \bar\psi_-\psi_+ \\
		=-2\Re \left(\bar\psi_+ \psi_{-{}}\right)\,.
\end{equation}
where in the last line we used $\left(\bar\psi_-\psi_+\right)^\dagger=-\bar\psi_+\psi_-$.
Using \eqref{psiUVABCD} we see that 
\begin{equation}
\label{E:QAgD}
	Q = -2\Re \left(A^{a\dagger} \gammat D^{a}\right). 
\end{equation}
Combining \eqref{E:QAgD} with \eqref{GotGRM}, it follows that
\eqn{GotQGr}{
Q = 2\Im \left(A^{\dagger}_{(b)} \gammat\, G^{\phantom{R}\,(a)}_{R\,\phantom{(a)}(b)} \gammat A^{(b)}\right)\,.
}
Since, $A$ is an arbitrary spinor and $\gammat$ is invertible, $Q<0$ implies that ${1\over 2i} \left( G - G^\dagger\right)$
is a nonpositive definite matrix and therefore, $\rho=-\Im \tr G$ is everywhere nonnegative.

It remains to show that with our IR asymptotics \eqref{psiIR}, $Q<0$.
To this end we work out the constraint on the $V_i$ that follows from \eqref{DCA}. If we split the $V_i$ into chiral spinors, denoted by $V_i^\pm$, the constraint can be written as
\eqn{DiracConstraintV}{
V_1^-= i {\alpha_+\over \kappa_+} V_1^+ \qquad V_2^-= i {\alpha_-\over \kappa_-} V_2^+
}
where
\eqn{alphapmDef}{
\alpha_{\pm} \equiv  -e^{\chi_{\rm IR}\over 2}\omega\, \gammat + {2 k_x  \pm \gYM W_{\rm IR} \over 2 c_{\rm IR}} \gamma^{\underline x} + k_y \gamma^{\underline y} \,.
}

There are now several different case we must consider. Suppose first that we are outside both IR Dirac cones. If we take the $V_i^+$ to be real and choose the Majorana basis for the $\gamma$ matrices (making them real matrices) it follows from \eqref{DiracConstraintV} that the $V_i^-$ are pure imaginary. Therefore, $\bar\psi_+\psi_-$ is also pure imaginary and $Q$ is identically zero leading us to conclude that $\rho=0$ (up to, perhaps, a codimension one surface of normal modes.) 
Next, suppose that $\kappa_+^2<0$ and take $V_2=0$ for simplicity. If we choose $V_1^+$ real then $V_1^-$ will also be real and $Q$ is no longer zero.  It is given by
\eqn{GotQInside}{
Q = -\frac{2 \sgn\omega}{|\kappa_+|} \Re\bigg( V_1^{+\dagger} \gammat\alpha_+ V_1^+\bigg)\,.
}
It is now a simple exercise to compute the eigenvalues of the matrix $\gammat \alpha_+$ and show that it is positive (negative) definite for $\omega>0$ ($\omega<0$). Therefore, $Q$ is always negative for $\kappa_+^2<0$ and $V_2=0$. We can show in the same way that $Q$ is negative for $\kappa_-^2<0$ and $V_1=0$. Since $Q$ is given by a quadratic form, it follows that $Q\geq 0$.

\section{Evaluating the spectral function}
\label{FERMIONS}

\subsection{Numerical results}
\label{NUMERICS}
To obtain the spectral function $\rho = - \Im \tr G^R$ we solved \eqref{psiEOM} numerically with the initial conditions given in \eqref{psiIR} for some small but nonzero $r_i$. We then extracted $A$ and $D$ from the behavior of the numerical solution at large $r$. This procedure was repeated for the four linearly independent choices of the $V_i$ in \eqref{psiIR}. We then used \eqref{Mdef} to obtain  $\mathcal M$ and \eqref{GotGRM} to obtain $G_{\rm R}$. To find the normal modes we looked for  zeros of $\det\, \mathcal{U}$. 

Let us now discuss our numerical results.  The spectral measure $\rho(k_m)$ is symmetric separately under the sign flips $\omega \to -\omega$, $k_x \to -k_x$, and $k_y \to -k_y$.  Since we are restricting our attention to massless fermions the only free parameter 
is $\gYM$, the Yang-Mills coupling constant. 
Most of  the features discussed in this section seem to be generic and independent of $\gYM$, with the obvious restriction that $\gYM$ is at least large enough so that the $T=0$ background we are considering exists. It seems possible that if we take $\gYM$ large enough there could be several disconnected surfaces of normal mode, as was found in \cite{Gubser:2009dt} for the $s$-wave background, but we were unable to check whether this happens as our numerics are not reliable for all of the ``preferred region'' for large $\gYM$. For $\gYM$ as large as 10, there is only one distinct surface. 

\begin{figure}
\begin{center}
\includegraphics[height = 7 cm]{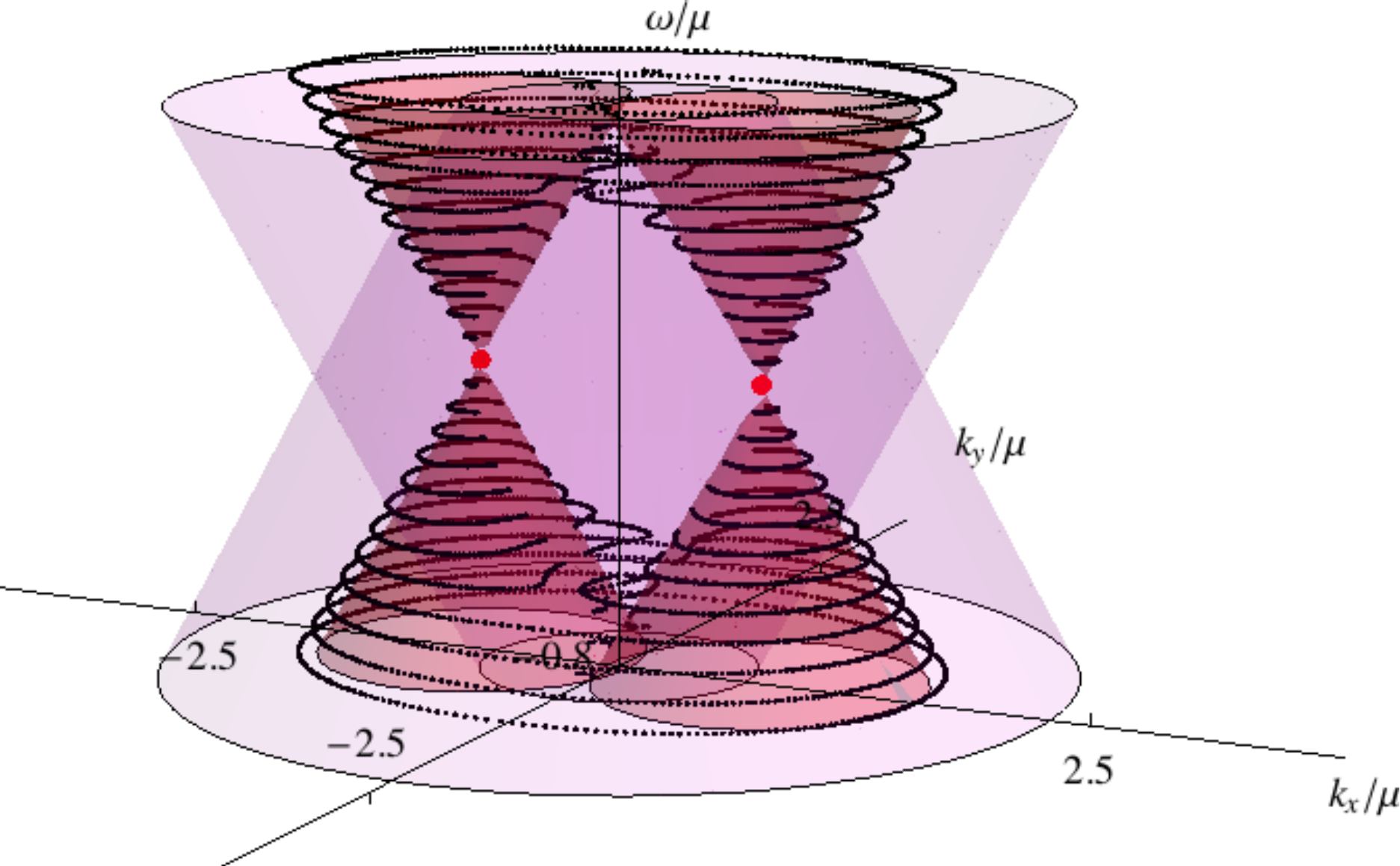} \\
\caption{The normal mode  surface for $\gYM=3$. The IR and UV Dirac cones are always drawn red and purple respectively. The normal mode surface is contained in the preferred region \eqref{GotRegion} and is shown as black dots. }
\label{FIG:NormalModesq3}
\end{center}
\end{figure}

The surface of normal modes we obtained numerically can be roughly described as a truncated deformed version of the IR Dirac cones, as can be  seen in  Fig.~\ref{FIG:NormalModesq3}.
This surface does not extend indefinitely and is contained in the preferred region \eqref{GotRegion}. 
It is perhaps more interesting to consider the shape of the normal mode surface near $\omega=0$. The reason for this is that, in our conventions where $\PHI(0)=0$, a fermion with $\omega=0$ has energy equal to the chemical potential.\footnote{If we wanted to treat $\Psi$ beyond the probe approximation, the natural construction is to fill all the excitations with energy below the chemical potential, i.e, all the normal modes with $\omega<0$. Once this is done, the Fermi surface will be given by the intersection of the normal mode surface with the $\omega=0$ plane.  But we don't understand how to systematically treat the back-reaction of the fermions on the geometry and the gauge field.}  For the $p$-wave superconductor, Fig.~\ref{FIG:NormalModesq3} clearly shows as $\omega \to 0$ the surface of normal modes approaches the IR Dirac cones and that therefore the Fermi surface is given by two isolated points: the apexes of the IR Dirac cones. 

\begin{figure}
\begin{center}
\includegraphics[height = 7 cm]{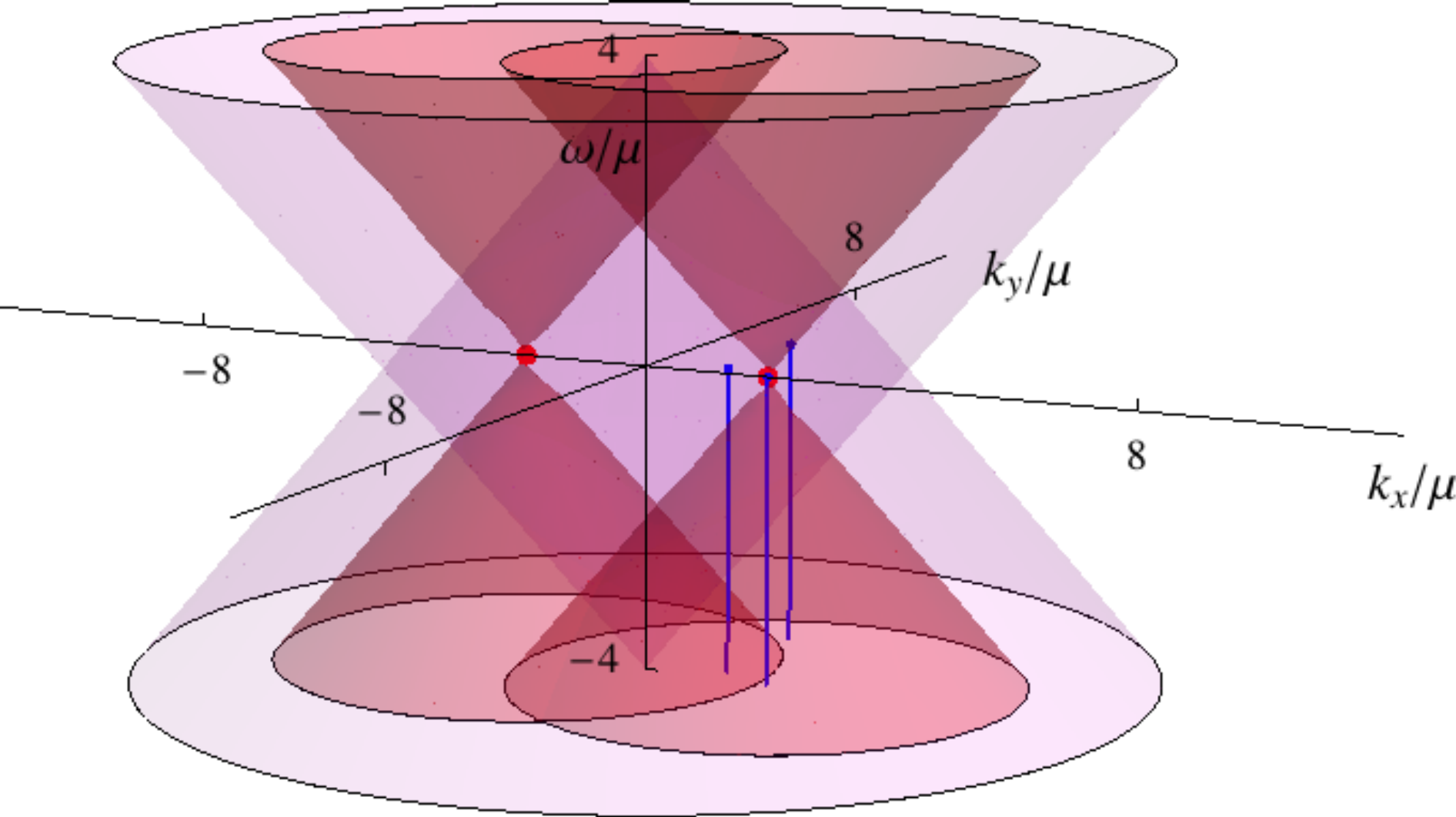} \\
\includegraphics[width = 6 in]{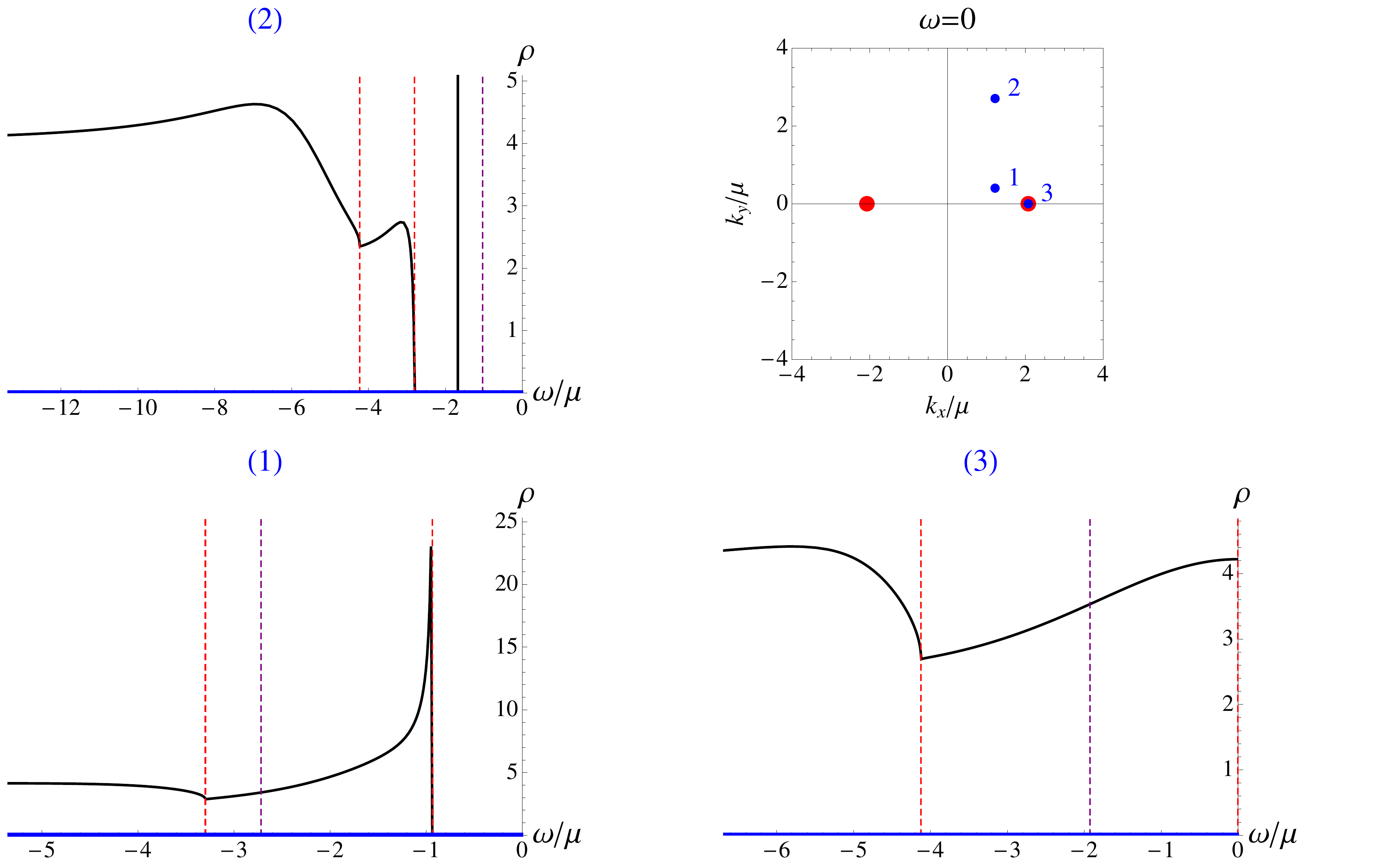}
\caption{\label{FIG:OmegaRho} The preferred region and the spectral measure $\rho$ as a function of $\omega$ for fixed $k_x,k_y$. The IR Dirac cones are shown in red, and UV Dirac cones in purple in the top figure. The blue lines are constant $k_x$ and $k_y$ lines and are the axes of the plots on the bottom. The limits of the IR and UV Dirac cones are marked in these plots as vertical dashed lines that are red and purple respectively. All the plots are for $\gYM=8$.}
\end{center}
\end{figure}

To understand the behavior of $\rho$ inside the IR Dirac cones it is useful to plot $\rho(k^m)$ as a function of $\omega$ for fixed $k_x,k_y$. Some of these plots are shown in Fig.~\ref{FIG:OmegaRho}. 
Outside the IR Dirac cones (the boundaries of which are shown as dashed red lines), $\rho$ vanishes except for, perhaps, a $\delta$-function peak that indicates the presence of a normal mode. Inside the IR Dirac cones, $\rho$ is positive and smooth except for the intersection of two Dirac cones where there is a kink.

For $k^m$ close to the apex of a Dirac cone, it is natural to expect that we will recover the behavior extracted from the infrared asymptotics.  This behavior was discussion in section~\ref{SUMMARY} for the $SO(4)$ case, and it is essentially the same in the $SU(2)$ case except in that there are only two Dirac cones rather than four.  It is clear from plot (3) in Fig.~\ref{FIG:OmegaRho} that $\rho$ goes to a constant as $\omega \to 0$ when $\vec{k}$ is at the apex of the Dirac cone.  Plot (1) in Fig.~\ref{FIG:OmegaRho} reveals that there is almost a power-law singularity as soon as one crosses into the Dirac cone close to its apex.  One does not see this dramatic near-singularity when crossing into a Dirac cone far from its apex.

\begin{figure}
\begin{center}
\includegraphics[height = 5.5 cm]{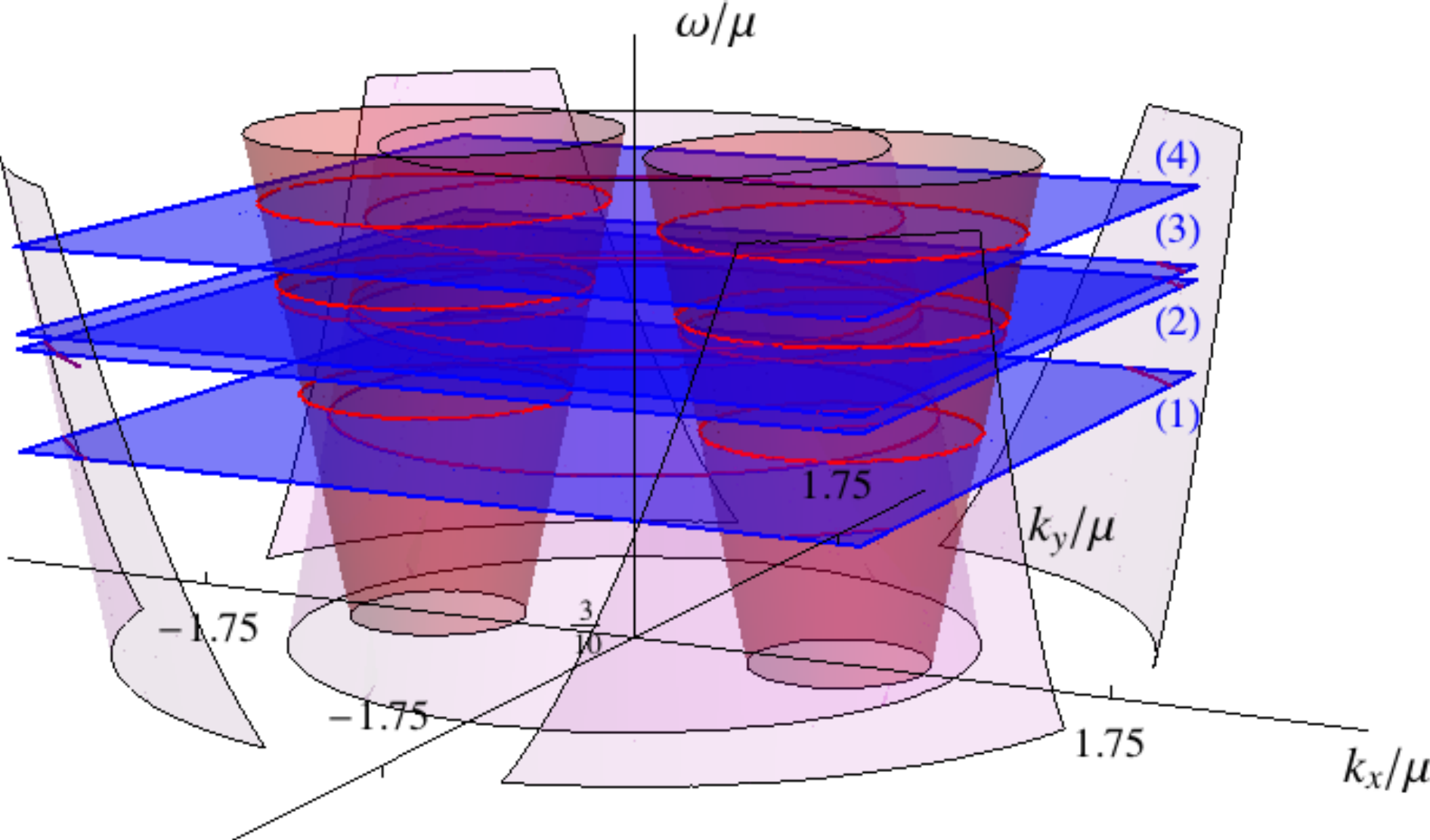} \\
\includegraphics[width = 5 in]{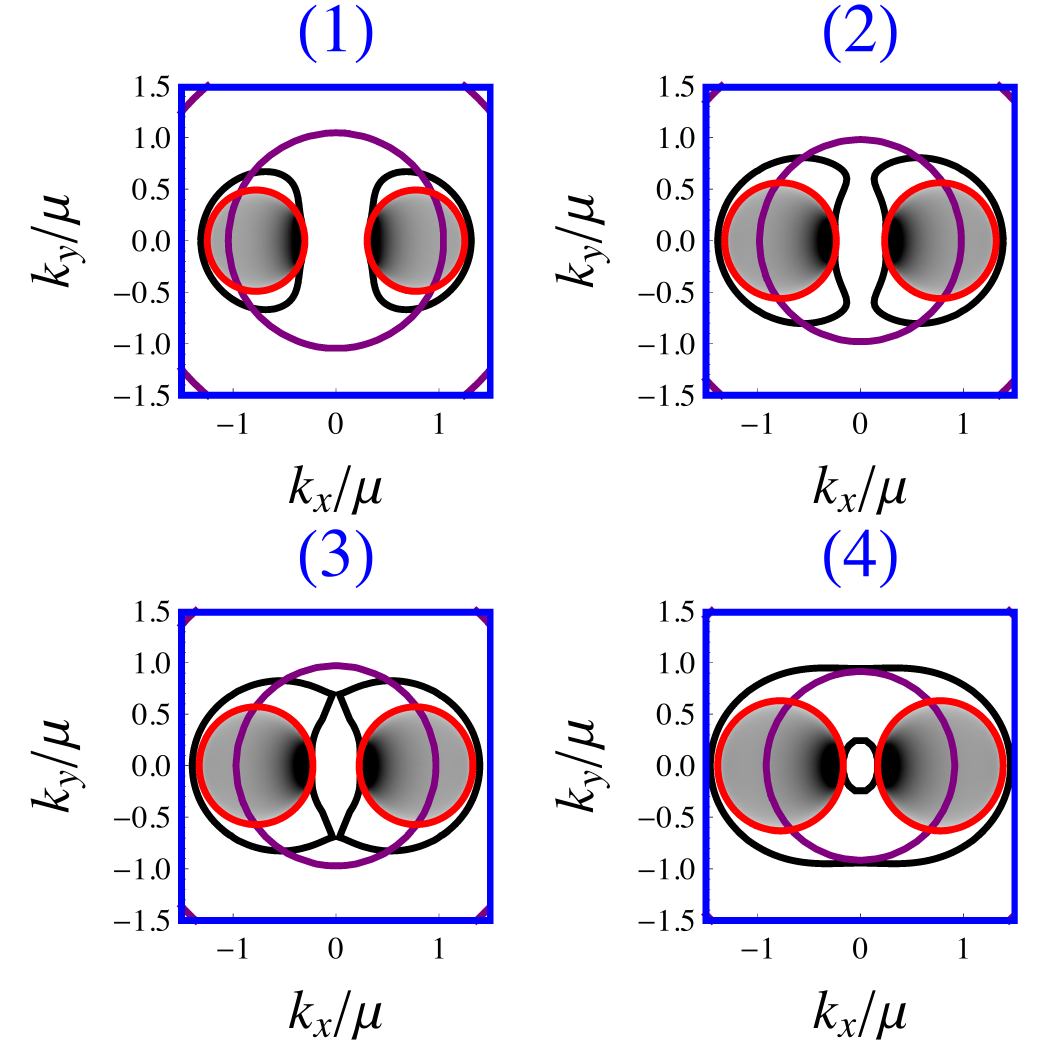}
\caption{The spectral measure $\rho$ for $\omega/\mu = 0.45,\,0.52,\,0.53,\,0.58$ and $\gYM=3$. The UV and IR Dirac cones are shown in the top figure as red and purple surfaces respectively. The blue planes are the constant $\omega$ planes for which $\rho$ is plotted in the bottom figure. In these plots, outside the the IR Dirac cones the surface of normal modes is shown as a black curve. This is overlaid with a density plot of the spectral measure $\rho$ where black corresponds to large values of $\rho$.}
\label{FIG:TouchSurface} 
\end{center}
\end{figure}

\begin{figure}
\begin{center}
\includegraphics[height = 6.5 cm]{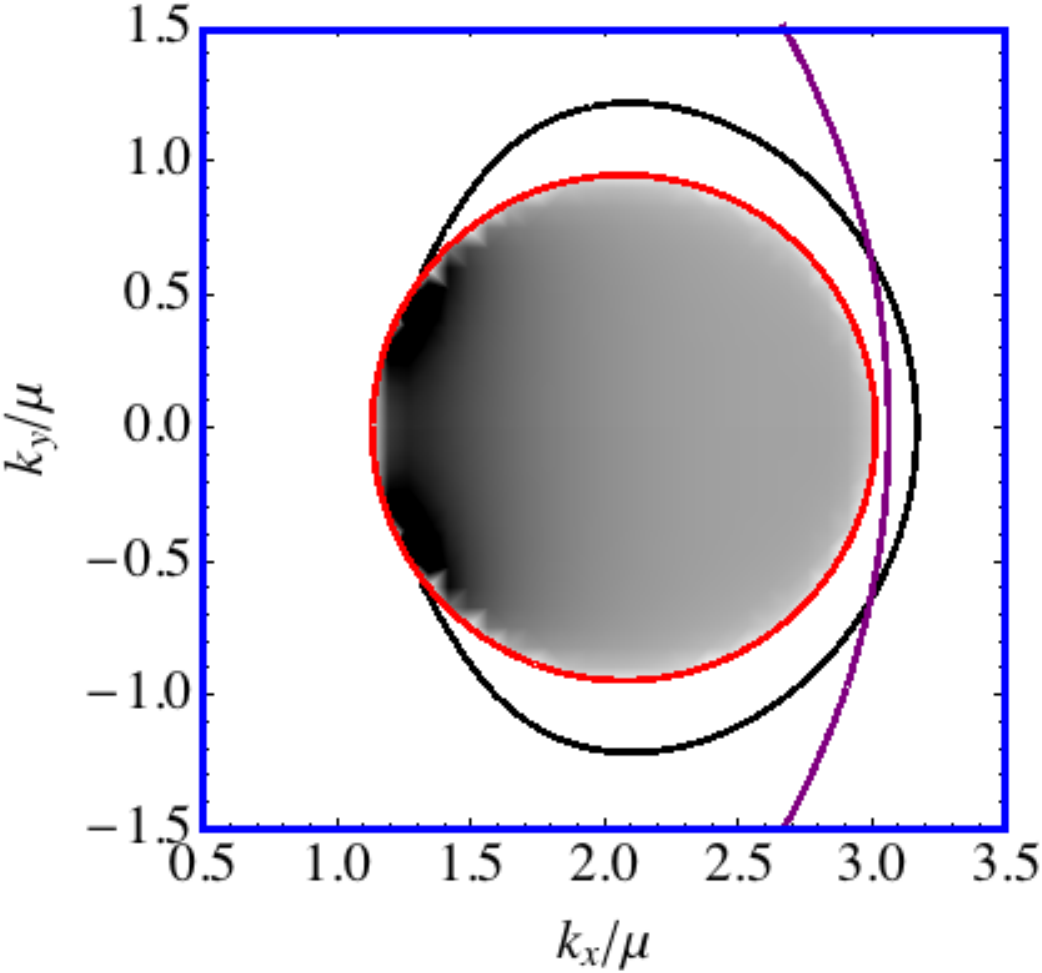}
\caption{A contour plot of the spectral function $\rho$ on a constant $\omega/\mu=0.93$ slice for $\gYM=8$. The red and purple curves mark the location of the UV and IR cones respectively. The black line outside the IR light cone shows the location of the normal modes. The spectral function $\rho$ develops a ridge between the points where the normal modes terminate on the light-cone.}
\label{FIG:Scar} 
\end{center}
\end{figure}

To gain a better understanding of the spectral measure we provide some plots of constant $\omega$ slices in figure~\ref{FIG:TouchSurface}. 
At small $\omega$ we find that the normal modes surround each of the IR light cones. For the particular case of $\gYM=3$ depicted in the top left corner of figure ~\ref{FIG:TouchSurface} the normal modes terminate on the light-cone leaving a ``scar'' in its interior. This is also seen more clearly in figure \ref{FIG:Scar}. 
This ridge inside the IR light-cone suggests that the normal mode has turned into a relatively long-lived quasinormal mode, or resonance.

Going back to figure \ref{FIG:TouchSurface}, we observe that as $\omega$ is increased the shape of the surface of normal modes becomes more asymmetric until, eventually, the surface of normal modes around each of the light cones intersect (figure \ref{FIG:TouchSurface} (3)). At large $\omega$ the normal modes arrange themselves into an inner an outer closed surface. This inner surface disappears once the IR Dirac cones start overlapping. The outer surface will also disappear, eventually, and this happens as soon as the outermost UV Dirac cone is inside the IR Dirac cones.

\subsection{Fermion correlators in the sudden approximation}
\label{PROBE}

The numerical work reported so far in this section turned up some interesting features in fermionic Green's functions: multiple thresholds associated with overlapping Dirac cones, normal modes in a preferred region, long-lived quasi-normal modes, and some signs of recovering the infrared approximations near the apex of a Dirac cone.  We would like, if possible, some analytic approximation that exposes these features more clearly, so that we are not so dependent upon numerical integration of classical equations of motion.  The obvious place to start is the probe approximation discussed in section~\ref{PROBEBGD}.  This approximation relies on having $\gYM$ big so that the gauge field doesn't back-react appreciably on the AdS${}_4$ geometry.  Recall that in section \ref{PROBEBGD} we rescaled fields as needed to eliminate explicit dependence of the equations of motion on $\gYM$, and that at $T=0$ we were able to replace the AdS${}_4$ geometry by the $z>0$ half of ${\bf R}^{3,1}$.  For the sake of simplicity we will continue to restrict to a massless Dirac fermion, which (after a suitable rescaling) we denote $\psi$.  It is dual to a spinorial operator ${\cal O}_\Psi$.  We do {\it not} restrict $\psi$ at this stage to be a doublet of $SU(2)$: it will become apparent that most of our discussion can be carried through straightforwardly for a domain wall configuration based on any semi-simple gauge group, with $\psi$ transforming in any representation of it.

The two-point functions of ${\cal O}_\Psi$ are controlled by solutions to the linear equation of motion for $\psi$:
 \eqn{MasslessDirac}{
  \Gamma^\mu (\partial_\mu - i A_\mu) \psi = 0 \,.
 }
Because we are working in the $z>0$ half of ${\bf R}^{3,1}$, we do not need to distinguish between $\Gamma^\mu$ and $\Gamma^{\underline\mu}$.  $A_\mu$ is a domain wall solution to the flat-space Yang-Mills equations, so for the $SU(2)$ case it is takes the form
 \eqn{AmForm}{
  SU(2):\qquad A_m dx^m = \Phi(z) \, dt \, \tau^3 + W(z) \, dx^1 \, \tau^1 \,,
 }
where $\Phi$ and $W$ satisfy the equations \eno{ProbeEOMs}.  For $SO(4)$, an interesting solution generalizing \eno{SOfourFlat} is
 \eqn{SOfourAm}{
  SO(4):\qquad A_m dx^m = \Phi(z) \, dt \, (\tau^3_A + \tau^3_B) + 
    W(z) \, (dx^1 \, \tau^1_A + dx^2 \, \tau^2_B) \,,
 }
with the same functions $\Phi(z)$ and $W(z)$ as in the $SO(4)$ case.  Backgrounds based on more complicated gauge groups could also be constructed.  All we require for the discussion of the next few paragraphs is that $A_m$ depends only on $z$, and $A_z = 0$ (a gauge choice).

Solutions to \eno{MasslessDirac} can be cast in the form
 \eqn{PsiForm}{
  \psi(x,z) = e^{i k \cdot x} \hat\psi(z) \,,
 }
where, as usual, $k_m = (-\omega,\vec{k})$ and $x^m = (t,\vec{x})$.  If we define
 \eqn{KmDefScaled}{
  K_m(z) = k_m - A_m(z) \,,
 }
then \eno{MasslessDirac} becomes
 \eqn{PsiHatEOM}{
  (\partial_z + i \Gamma^z \Gamma^m K_m) \hat\psi = 0 \,,
 }
whose solutions can be formally expressed as
 \eqn{FormalSolns}{
  \hat\psi(z) = P \left\{ \exp\left[ -i \int_0^z dz' \, \Gamma^z \Gamma^m K_m(z') \right] \right\}
    \hat\psi(0) \,,
 }
where $P$ denotes ordering non-commuting matrices so that those coming from larger values of $z$ go to the left.  The allowed solutions satisfy boundary conditions in the infrared (large $z$) which can be compactly expressed as
 \eqn{IRbc}{
  \hat\psi \propto e^{-K_{\rm IR} z} u \,,
 }
where restrictions on $u$ come only from solving \eno{PsiHatEOM}, and we define
 \eqn{LimDefs}{\seqalign{\span\TL & \span\TR &\qquad\span\TL & \span\TR}{
  K_m^{\rm IR} &= \lim_{z \to \infty} K_m(z) & K_{\rm IR} &= \sqrt{K^m_{\rm IR} K_m^{\rm IR}}  \cr
  K_m^{\rm UV} &= \lim_{z \to \infty} K_m(z) & K_{\rm UV} &= \sqrt{K^m_{\rm UV} K_m^{\rm UV}} \,,
 }}
and the $i\epsilon$ prescription \eno{PolePassing} is implied.

 \begin{figure}
  \centerline{\includegraphics[width=4in]{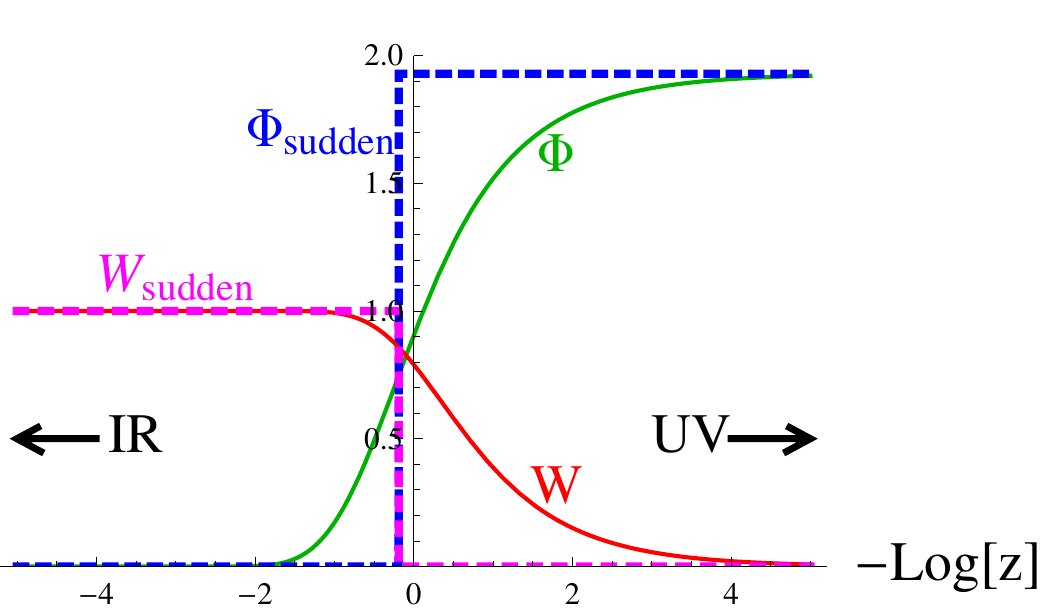}}
  \caption{The functions $\Phi$ and $W$, with $W_{\rm IR}=1$, together with the sudden approximations $\Phi_{\rm sudden}$ and $W_{\rm sudden}$ discussed in the main text. }\label{FIG:SuddenApprox}
 \end{figure}
It is hard to go further than \eno{FormalSolns} without some additional approximation because the path-ordered exponential is hard to compute.  The domain wall structure suggests an obvious approximation, illustrated in figure~\ref{FIG:SuddenApprox}: Let's replace $A_m \to A_m^{\rm sudden}$, where $A_m^{\rm sudden}$ is piecewise constant and piecewise flat, going straight from the UV flat connection to the IR flat connection at a special value $z_*$ of $z$ such that
 \eqn{DetermineZstar}{
  \int_0^{z_*} dz \, A_t^{\rm sudden} = \int_0^\infty dz \, A_t \,.
 }
A straightforward calculation based on the function $\tilde\Phi(\zeta)$ introduced in section~\ref{PROBEBGD} shows that in the $SU(2)$ and $SO(4)$ cases described by \eno{AmForm} and \eno{SOfourAm},
 \eqn{zStarValue}{
  W_{\rm IR} z_* \approx 1.2058 \,.
 }
From now on we will set $W_{\rm IR} = 1$ for simplicity.  It is easy to solve \eno{MasslessDirac} with $A_m$ replaced by $A_m^{\rm sudden}$: the solution is
 \eqn{PsiSudden}{
  \hat\psi_{\rm sudden}(z) = \left\{ \seqalign{\span\TL &\qquad\span\TT}{
    e^{-iz \Gamma^z \Gamma^m K_m^{\rm UV}} \hat\psi_{\rm sudden}(0) & for $0 < z < z_*$  \cr
    e^{-i(z-z_*) \Gamma^z \Gamma^m K_m^{\rm IR}} e^{-iz_* \Gamma^z \Gamma^m K_m^{\rm UV}}
      \hat\psi_{\rm sudden}(0) & for $z_* < z \,.$
    } \right.
 }
Now let's see how to implement the boundary condition \eno{IRbc}.  Because $\hat\psi_{\rm sudden}$ satisfies the Dirac equation (with the replacement $A_m \to A_m^{\rm sudden}$), we have
 \eqn{OneHand}{
  \partial_z \hat\psi_{\rm sudden}(z) = 
    -i \Gamma^z \Gamma^m K_m^{\rm IR} \hat\psi_{\rm sudden}(z) \qquad\hbox{for all $z>z_*$.}
 }
On the other hand, $\partial_z \hat\psi_{\rm sudden}(z) = -K_{\rm IR} \hat\psi_{\rm sudden}(z)$ for large $z$ because of \eno{IRbc}.  Since the $K_m^{\rm IR}$ commute with one another, it must be that
 \eqn{OtherHand}{
  \partial_z \hat\psi_{\rm sudden}(z) = -K_{\rm IR} \hat\psi_{\rm sudden}(z) \qquad\hbox{for all $z>z_* \,.$}
 }
Comparing \eno{OneHand} and \eno{OtherHand}, we arrive at
 \eqn{ConstrainPsiZero}{
  P \hat\psi_{\rm sudden}(0) = 0
 }
where
 \eqn{DefineP}{
  P \equiv (K_{\rm IR} - i \Gamma^z \Gamma^m K_m^{\rm IR}) e^{-i z_* \Gamma^z \Gamma^n K_n^{\rm UV}} 
     \,.
 }
It will be useful to express
 \eqn{ExpressP}{
  P = q + \Gamma^z \Gamma^m q_m + \Gamma^m \Gamma^n q_{mn}
 }
where
 \eqn{qDefs}{
  q &= K_{\rm IR} \cosh(z_* K_{\rm UV})  \cr
  q_m &= -i \left[ K_m^{\rm IR} \cosh(z_* K_{\rm UV}) + K_{\rm IR} K_m^{\rm UV} 
    {\sinh(z_* K_{\rm UV}) \over K_{\rm UV}} \right]  \cr
  q_{mn} &= K_m^{\rm IR} K_n^{\rm UV} {\sinh(z_* K_{\rm UV}) \over K_{\rm UV}} \,.
 }
Note that $q$, $q_m$, and $q_{mn}$ have no spinor structure: they are purely gauge-theoretic quantities.  In \eno{ExpressP}, the first term on the right hand side is implicitly multiplied by the unit matrix in spinor space. 

The equation \eno{ConstrainPsiZero} determines the fermionic Green's function in the sudden approximation, as we now explain.  If we use our usual basis for gamma matrices, \eno{GammaExpress}, and express
 \eqn{ChiralExpress}{
  \hat\psi_{\rm sudden}(0) = \begin{pmatrix} u_+ \\ u_- \end{pmatrix} \,,
 }
then \eno{ConstrainPsiZero} can be recast as
 \eqn{AbstractConstraint}{
  \begin{pmatrix} P_{++} & P_{+-} \\ P_{-+} & P_{--} \end{pmatrix}
   \begin{pmatrix} u_+ \\ u_- \end{pmatrix} = 0 \,,
 }
where
 \eqn{FoundPpm}{
  \begin{pmatrix} P_{++} & P_{+-} \\ P_{-+} & P_{--} \end{pmatrix} = 
   \begin{pmatrix} q + \gamma^m \gamma^n q_{mn} & -\gamma^m q_m  \\
     \gamma^m q_m & q + \gamma^m \gamma^n q_{mn} \end{pmatrix} \,.
 }
Comparing to \eno{psiUVABCD} and using \eno{Mdef}, we arrive at
 \eqn{SuddenG}{
  G_{\rm sudden}(k) = i P_{+-}^{-1} P_{++} \gammat = 
    i P_{--}^{-1} P_{-+} \gammat \,.
 }
A more explicit form, based on the middle expression in \eno{SuddenG}, is
 \eqn{SuddenGAgain}{
  G_{\rm sudden}(k) = -i (\gamma^m q_m)^{-1} (q + \gamma^m \gamma^n q_{mn}) \gammat \,.
 }
We further define $\rho_{\rm sudden}(k) \equiv -\Im \tr G^R_{\rm sudden}(k)$.  We note that $\tr G^R_{\rm sudden}(k)$ and hence $\rho_{\rm sudden}(k)$ can in principle be found in closed form as functions of $k_m$, $\mu$, $W_{\rm IR}$, and $z_*$.  In practice, the closed-form expressions for $q$, $q_m$, and $q_{mn}$ in terms of $k_m$, $\mu$, $W_{\rm IR}$, and $z_*$ are already quite complicated, and we were unable to find a closed-form expression for the inverse matrix $(\gamma^m q_m)^{-1}$ that was sufficiently compact to be useful. 

Although we did not succeed in finding a simple enough closed form expression for the spectral measure to record here, the expression \eno{SuddenGAgain} is simple enough to expose most of the qualitative features of the analytic structure of $G_{\rm sudden}(k)$.  First we claim that branch cuts in $q$ and $q_m$, as functions of $k_m$, arise only when $K_{\rm IR}$ has a branch cut, while $q_{mn}$ has no branch cuts at all.  To demonstrate this claim, we observe that $\cosh(z_*K_{\rm UV})$ and ${\sinh(z_*K_{\rm UV}) \over K_{\rm UV}}$ are analytic functions of $K_{\rm UV}^2$, which in turn is a quadratic expression in the momenta $k_m$; so $\cosh(z_*K_{\rm UV})$ and ${\sinh(z_*K_{\rm UV}) \over K_{\rm UV}}$ have no branch cuts at all as functions of the $k_m$.  More trivially, $K_m^{\rm IR}$ and $K_m^{\rm UV}$ also have no branch cuts.  Our first claim now follows by inspection of the formulas \eno{qDefs} for $q$, $q_m$, and $q_{mn}$.

It follows from our first claim that the continuum part of $\rho_{\rm sudden}(k)$ is supported precisely where $K_{\rm IR}$ has a branch cut, which is to say inside the Dirac cones.  This feature of the spectral function has been discussed extensively in section \ref{SPECTRAL} and is essentially as expected on intuitive grounds: The retarded Green's function can have a dissipative part iff some component of the fermion wave-function is infalling in the infrared, rather than exponentially decaying there.

Our second claim is that, at least for $\psi$ in a real representations of ${\bf g}$ and for generic values of $\vec{k} = (k_x,k_y)$, $\delta$-function singularities in $\rho_{\rm sudden}(\omega,k_x,k_y)$ as a function of $\omega$ can only occur outside the IR Dirac cones.  Recall that a $\delta$-function in $\rho_{\rm sudden}(\omega,k_x,k_y)$ is associated with a pole in $\Re \tr G_{\rm sudden}^R(\omega,k_x,k_y)$ for real $\omega$.  Thus our claim is that any such pole must arise outside the Dirac cones.  To demonstrate the claim, first note that $q$, $q_m$, and $q_{mn}$ never diverge.  So the only way to get a pole is if $\gamma^m q_m$ is non-invertible, which is to say $\det i \gamma^m q_m$ vanishes.  The $\gamma^m$ may be chosen in a Majorana basis, where all entries are real: indeed, \eno{GammaChoice} is such a basis.  In a real representation of the gauge group, all the $K_m^{\rm IR}$ and $K_m^{\rm UV}$ are real symmetric matrices, as are $\cosh(z_*K_{\rm UV})$ and ${\sinh(z_*K_{\rm UV}) \over K_{\rm UV}}$.  $K_{\rm IR}$ is also real and symmetric provided we are outside the Dirac cones.  Thus the $iq_m$ are real matrices, and we see that $\det i \gamma^m q_m$ must indeed be real.  More precisely, for real $k_x$ and $k_y$, $\det i \gamma^m q_m$ is a real function of the variable $\omega$ outside the Dirac cones.  Inside the Dirac cones, $K_{\rm IR}$ has an imaginary part, and $\det i \gamma^m q_m$ is {\it not} a real function of $\omega$.  All that we need to note now in order to complete our argument is that real analytic functions of a single variable generically can have zeros on the real axis, but general complex analytic functions do not.  It is tempting to speculate that this argument could be extended to fermions in complex representations.  Certainly there is intuitive reason to think that when a delta-function contribution to the spectral measure crosses into a continuum, it will spread out into a finite-width resonance---as we saw numerically in Fig.~\ref{FIG:Scar}.

We caution the reader that the sudden approximation is {\it not} controlled in the sense of becoming good when some parameter is taken large or small.  Usually, sudden approximations are justified when wave-functions are slowly varying as compared to the features of the underlying background that one is approximating.  Optimistically, one might expect our sudden approximation to be good near the infrared light-cone, because then the fermion wave-functions are slowly varying in the region $z>z_*$.  But these wave-functions are not necessarily slowly varying for $z<z_*$.  Thus we regard \eno{SuddenGAgain} as useful in the sense of providing an in-principle closed-form expression that captures some of the relevant physics: namely a continuous part of the spectral measure inside the Dirac cones, with the possibility (at least on genericity grounds) of normal modes only outside the cones.

\section{Discussion}
\label{DISCUSSION}

The starting point of our analysis is the classical action
 \eqn{GravityPlusGauge}{
  S = \int_M d^4 x \sqrt{-g} \left( R + {6 \over L^2} - {1 \over 2} \tr F_{\mu\nu}^2 - 
    i \bar\Psi \Gamma^\mu D_\mu \Psi \right) + \hbox{boundary terms} \,,
 }
which is essentially the lagrangian of QCD coupled to gravity with a negative cosmological constant, except that we choose the gauge group to be $SU(2)$ or $SO(4)$, while the fermion transforms either as the doublet of $SU(2)$ or the fundamental ${\bf 4}$ of $SO(4)$.  The lagrangian \eno{GravityPlusGauge} describes the bulk dynamics dual to a field theory in $2+1$ dimensions whose continuous symmetries form the same group as the gauge group in \eno{GravityPlusGauge}.  We treat the dynamics of \eno{GravityPlusGauge} classically, which is understood to be dual to a large $N$ approximation in the field theory.

One output of our analysis is the phase diagram of superconducting black holes based on the $SU(2)$ gauge group.  This phase diagram is shown in Fig.~\ref{F:PTtotal}.  We demonstrated, largely through a numerical study, that black holes charged under the $\tau^3$ generator of $SU(2)$ spontaneously break that symmetry through a $p$-wave condensate similar to the one originally studied in \cite{Gubser:2008wv}.  A conspicuous feature of the phase diagram is a tricritical point separating second order and first order behavior at the symmetry breaking phase transition.  We also showed that at low temperatures, the symmetry-breaking solutions approach the AdS${}_4$-to-AdS${}_4$ domain wall geometries of \cite{Basu:2009vv}, similar to domain walls found in the Abelian Higgs model in \cite{Gubser:2008wz} except for anisotropic alteration of the coordinate speed of light in the infrared.  Our analysis is not complete in that we did not systematically study the stability of the symmetry-breaking solutions, and it is possible that there are other symmetry-breaking configurations that we missed.  Thus we cannot rule out the existence of a more complicated phase diagram than we plotted in Fig.~\ref{F:PTtotal}, with (for example) symmetry-breaking phases present below $\gYM = 0.710$.

A second output of our analysis is two-point functions of the fermionic operators dual to $\Psi$.  These two-point functions show some intriguing parallels with ARPES data on high-temperature superconductors.  Relationships between holographic fermionic correlators and ARPES data were emphasized early in \cite{Cubrovic:2009ye} in the context of the non-superconducting phase, following earlier work \cite{Lee:2008xf,Iqbal:2009fd,Liu:2009dm}.  Studies in the superconducting phase include \cite{Faulkner:2009am,Chen:2009pt}; see also \cite{Gubser:2009dt}.  These works all focused on rotationally symmetric backgrounds.  By contrast, our Fermi surface at $T=0$ consists of isolated points: two in the $SU(2)$ example, and four in the $SO(4)$ example.  Above each isolated point, a Dirac cone rises, as can be seen in Fig.~\ref{FIG:OmegaRho} for the $SU(2)$ case and in Fig.~\ref{IRPLOTS2} for the $SO(4)$ case.  In the $SO(4)$ case, the nodes are positioned at $45^\circ$ degrees relative to the axes along which the gauge potentials are aligned, reminding us of the positioning of nodes in the gap in $d_{x^2-y^2}$ superconductors.  

The structure of normal modes is also favorable to the comparison with ARPES data.  As shown in Fig.~\ref{FIG:NormalModesq3}, there is a normal mode slightly outside each Dirac cone.  And as shown in Fig.~\ref{FIG:OmegaRho} (plot (2) especially), the spectral measure significantly away from the tip of the Dirac cone exhibits a peak-dip-hump structure.  The peak comes from the normal mode, which shows up in the spectral measure at $T=0$ as a $\delta$-function, like an infinitely sharp quasi-particle.  The hump comes from the continuum part of the spectral measure, which is entirely inside the Dirac lightcone.  At the tip of the light cone (plot (2) of Fig.~\ref{FIG:OmegaRho}) or just slightly away from it (plot (1)), there is less structure: the peak goes away or merges into the hump.  Again we see a point of comparison with ARPES: the classic peak-dip-hump structure arises away from the node in the gap.

There is a twist in our discussion of Dirac cones, normal modes, and continuum structures relative to the usual story based on quasi-particles, where continuum structures arise because of two- or three-particle states, where each particle by itself is on-shell when its momentum lies on the Dirac cone.  In our case, the Dirac cone characterizes the edge of the continuum rather than the dispersion relation for the quasi-particle.  We see already from formulas like \eno{GotGPsi} that if only the infrared dynamics are accounted for, a continuum supported inside each Dirac cone is the {\it only} feature of the spectral measure.  There are no quasi-particles in sight in this infrared limit.  The quasi-particles (or, at least, the normal modes of the bulk fermions) come from the more intricate domain wall structure of the full bulk geometry, as discussed in section~\ref{NORMAL}.  The dispersion relation of these quasi-particles is not perfectly linear.  This is striking because, even in the full domain wall geometry, the continuum part of the spectral measure is supported in Dirac cones which are perfectly linear.  In fact, close inspection of Fig.~\ref{FIG:NormalModesq3} shows that the normal modes cross into the Dirac cones in certain regions.  This behavior is brought out better in Figs.~\ref{FIG:TouchSurface} and~\ref{FIG:Scar}.  Moreover, the normal modes disappear altogether once one passes outside the preferred region described in \eno{GotRegion}.  So for large enough $\omega$ and $\vec{k}$ there are no normal modes, and therefore no well-defined quasi-particle $\delta$-functions in the spectral measure.  But one still finds that the edge of the continuous part of the spectral measure defines a perfect Dirac cone.  In short, the continuum part of the spectral measure is the fundamental feature, while the quasi-particle $\delta$-function appears only under the right circumstances.

We were able to produce an explicit formula \eno{SuddenGAgain} which captures the main features of the fermionic Green's function, including the perfect Dirac cones enclosing the continuous part of the spectral measure and the normal mode outside the Dirac cones.  Previous works, notably \cite{Faulkner:2009wj}, have provided analytic approximations to interesting fermionic correlators; moreover, the ones in \cite{Faulkner:2009wj} are based on a controlled approximation, whereas ours is not.  The analytic forms found in \cite{Faulkner:2009wj} for the non-superconducting state rely upon the existence of an $AdS_2$ near-horizon region.  This feature of the geometry is double-edged: While it does provide tractable asymptotics, it also forces the existence of non-zero entropy density at zero temperature, which remains unexplained and seems to us peculiar in a theory whose underlying formulation is a continuum field theory rather than a lattice.  Limited analytic information about fermionic two-point functions is available in the backgrounds studied in \cite{Gubser:2009qt}, where also entropy vanishes linearly with temperature.  A more powerful understanding might be forthcoming if one better exploited the $AdS_3$ near-horizon region of the ten-dimensional embedding of these backgrounds.

There are some good reasons to be suspicious of the relevance of our setup to real high-$T_c$ materials with a $d$-wave gap:
 \begin{itemize}
  \item As already remarked in the introduction, the field theory dual to the $AdS_4$ bulk is a large $N$ field theory formulated in the continuum rather than on the lattice.
  \item The condensate in the $SO(4)$ case is not described in terms of a spin-$2$ bulk field, as one might expect, but rather in terms of gauge potentials involving off-diagonal generators of $SO(4)$.  It's not at all clear that the classic phase-sensitive features of the $d$-wave gap would show up in our system.
  \item Off-diagonal gauge potentials are dual to persistent currents of global symmetries in the boundary theory.  This seems rather different from the usual language for discussing $d$-wave superconductivity.  It would be interesting to see if the notion of persistent currents as an order parameter might be translated into lattice language, and how it might interact with constraints such as those discussed in \cite{Bloch:1968ab}.
  \item In addition to the $SO(4)$ gauge symmetry, the field theory dual to \eno{GravityPlusGauge} has $SO(2,1)$ Lorentz symmetry and also relativistic conformal invariance.  These symmetries are largely broken by the condensates, but the fermion operator ${\cal O}_\Psi$ transforms as a doublet under $SO(2,1)$.  This has no immediate analog in spin systems relevant to high-$T_c$ materials, where the spins are doublets under the $SU(2)_{\rm spin}$ and have no further structure under the Lorentz group in $2+1$ dimensions.
  \item The $SO(4)$ symmetry of the Hubbard model is composed of $SU(2)_{\rm spin}$ and $SU(2)_{\rm pseudospin}$.  The definition of the latter seems to require the lattice.  Doping amounts to adding a chemical potential for the $\tau^3$ component of pseudospin, whereas in \eno{SOfourAm} we added equal chemical potentials for the $\tau^3$ components of both $SU(2)$'s in $SO(4)$.
 \end{itemize}

Nevertheless, the resemblance of our results for the spectral measure of fermionic Green's functions to the spectral properties revealed in real materials by ARPES are striking enough that we should inquire what underlying physics is driving it.  At one level there is no puzzle: as soon as we note that the infrared Green's functions depend on Lorentz-invariant combinations of $K_m = k_m - \gYM A_m$, where $k_m = (-\omega,\vec{k})$ and $A_m$ is a flat connection in the gauge group of \eno{GravityPlusGauge}, we see that the displaced Dirac cones are just a consequence of the Lorentz invariance plus the eigenvalues of the non-zero components of $A_m$.  At another level, it may seem strangely suggestive that $SO(4)$ is the symmetry group of the Hubbard model on a bipartite lattice, and the fermion creation and annihilation operators transform as the ${\bf 4}$ of $SO(4)$.\footnote{A theory of phase competition between anti-ferromagnetic order and superconductivity has been advanced based on an approximate $SO(5)$ symmetry \cite{Zhang:1997so}; for a review see \cite{RevModPhys.76.909}.}  Precisely this choice of gauge group and fermion representation gave us the Dirac cone structure reminiscent of $d_{x^2-y^2}$ pairing.  Did we get approximately right answers for ARPES-like spectra because we have captured some correct features of the Hubbard model?  A positive answer to this question would be fairly exciting.

\section*{Acknowledgments}

We thank D.~Haldane, D.~Huse, C.~Mathy, and S.~Sondhi for useful discussions. This work was supported in part by the Department of Energy under Grant No. DE-FG02-91ER40671 and by the NSF under award number PHY-0652782. FDR was also supported in part by the FCT grant SFRH/BD/30374/2006.

\bibliographystyle{JHEP}
\bibliography{arc}

\providecommand{\href}[2]{#2}\begingroup\raggedright\begin{thebibliography}{10}

\bibitem{Gubser:2008px}
S.~S. Gubser, {\it {Breaking an Abelian gauge symmetry near a black hole
  horizon}},  {\em Phys. Rev.} {\bf D78} (2008) 065034,
  [\href{http://xxx.lanl.gov/abs/0801.2977}{{\tt arXiv:0801.2977}}].

\bibitem{Hartnoll:2008vx}
S.~A. Hartnoll, C.~P. Herzog, and G.~T. Horowitz, {\it {Building a Holographic
  Superconductor}},  {\em Phys. Rev. Lett.} {\bf 101} (2008) 031601,
  [\href{http://xxx.lanl.gov/abs/0803.3295}{{\tt arXiv:0803.3295}}].

\bibitem{Gubser:2008zu}
S.~S. Gubser, {\it {Colorful horizons with charge in anti-de Sitter space}},
  {\em Phys. Rev. Lett.} {\bf 101} (2008) 191601,
  [\href{http://xxx.lanl.gov/abs/0803.3483}{{\tt arXiv:0803.3483}}].

\bibitem{Gubser:2008wv}
S.~S. Gubser and S.~S. Pufu, {\it {The gravity dual of a p-wave
  superconductor}},  {\em JHEP} {\bf 11} (2008) 033,
  [\href{http://xxx.lanl.gov/abs/0805.2960}{{\tt arXiv:0805.2960}}].

\bibitem{Roberts:2008ns}
M.~M. Roberts and S.~A. Hartnoll, {\it {Pseudogap and time reversal breaking in
  a holographic superconductor}},  {\em JHEP} {\bf 08} (2008) 035,
  [\href{http://xxx.lanl.gov/abs/0805.3898}{{\tt arXiv:0805.3898}}].

\bibitem{Franco:2009yz}
S.~Franco, A.~Garcia-Garcia, and D.~Rodriguez-Gomez, {\it {A general class of
  holographic superconductors}},  \href{http://xxx.lanl.gov/abs/0906.1214}{{\tt
  arXiv:0906.1214}}.

\bibitem{Aprile:2009ai}
F.~Aprile and J.~G. Russo, {\it {Models of Holographic superconductivity}},
  \href{http://xxx.lanl.gov/abs/0912.0480}{{\tt arXiv:0912.0480}}.

\bibitem{Franco:2009if}
S.~Franco, A.~M. Garcia-Garcia, and D.~Rodriguez-Gomez, {\it {A holographic
  approach to phase transitions}},  {\em Phys. Rev.} {\bf D81} (2010) 041901,
  [\href{http://xxx.lanl.gov/abs/0911.1354}{{\tt arXiv:0911.1354}}].

\bibitem{Hartnoll:2009sz}
S.~A. Hartnoll, {\it {Lectures on holographic methods for condensed matter
  physics}},  {\em Class. Quant. Grav.} {\bf 26} (2009) 224002,
  [\href{http://xxx.lanl.gov/abs/0903.3246}{{\tt arXiv:0903.3246}}].

\bibitem{Herzog:2009xv}
C.~P. Herzog, {\it {Lectures on Holographic Superfluidity and
  Superconductivity}},  {\em J. Phys.} {\bf A42} (2009) 343001,
  [\href{http://xxx.lanl.gov/abs/0904.1975}{{\tt arXiv:0904.1975}}].

\bibitem{Horowitz:2010gk}
G.~T. Horowitz, {\it {Introduction to Holographic Superconductors}},
  \href{http://xxx.lanl.gov/abs/1002.1722}{{\tt arXiv:1002.1722}}.

\bibitem{RevModPhys.75.473}
A.~Damascelli, Z.~Hussain, and Z.-X. Shen, {\it Angle-resolved photoemission
  studies of the cuprate superconductors},  {\em Rev. Mod. Phys.} {\bf 75}
  (Apr, 2003) 473--541.

\bibitem{Chen:2009pt}
J.-W. Chen, Y.-J. Kao, and W.-Y. Wen, {\it {Peak-Dip-Hump from Holographic
  Superconductivity}},  \href{http://xxx.lanl.gov/abs/0911.2821}{{\tt
  arXiv:0911.2821}}.

\bibitem{Faulkner:2009am}
T.~Faulkner, G.~T. Horowitz, J.~McGreevy, M.~M. Roberts, and D.~Vegh, {\it
  {Photoemission 'experiments' on holographic superconductors}},
  \href{http://xxx.lanl.gov/abs/0911.3402}{{\tt arXiv:0911.3402}}.

\bibitem{Gubser:2009dt}
S.~S. Gubser, F.~D. Rocha, and P.~Talavera, {\it {Normalizable fermion modes in
  a holographic superconductor}},
  \href{http://xxx.lanl.gov/abs/0911.3632}{{\tt arXiv:0911.3632}}.

\bibitem{Lee:2008xf}
S.-S. Lee, {\it {A Non-Fermi Liquid from a Charged Black Hole: A Critical Fermi
  Ball}},  {\em Phys. Rev.} {\bf D79} (2009) 086006,
  [\href{http://xxx.lanl.gov/abs/0809.3402}{{\tt arXiv:0809.3402}}].

\bibitem{Cubrovic:2009ye}
M.~Cubrovic, J.~Zaanen, and K.~Schalm, {\it {String Theory, Quantum Phase
  Transitions and the Emergent Fermi-Liquid}},  {\em Science} {\bf 325} (2009)
  439--444, [\href{http://xxx.lanl.gov/abs/0904.1993}{{\tt arXiv:0904.1993}}].

\bibitem{Liu:2009dm}
H.~Liu, J.~McGreevy, and D.~Vegh, {\it {Non-Fermi liquids from holography}},
  \href{http://xxx.lanl.gov/abs/0903.2477}{{\tt arXiv:0903.2477}}.

\bibitem{Basu:2009vv}
P.~Basu, J.~He, A.~Mukherjee, and H.-H. Shieh, {\it {Hard-gapped Holographic
  Superconductors}},  \href{http://xxx.lanl.gov/abs/0911.4999}{{\tt
  arXiv:0911.4999}}.

\bibitem{Basu:2008bh}
P.~Basu, J.~He, A.~Mukherjee, and H.-H. Shieh, {\it {Superconductivity from
  D3/D7: Holographic Pion Superfluid}},  {\em JHEP} {\bf 11} (2009) 070,
  [\href{http://xxx.lanl.gov/abs/0810.3970}{{\tt arXiv:0810.3970}}].

\bibitem{Herzog:2009ci}
C.~P. Herzog and S.~S. Pufu, {\it {The Second Sound of SU(2)}},  {\em JHEP}
  {\bf 04} (2009) 126, [\href{http://xxx.lanl.gov/abs/0902.0409}{{\tt
  arXiv:0902.0409}}].

\bibitem{Ammon:2009xh}
M.~Ammon, J.~Erdmenger, V.~Grass, P.~Kerner, and A.~O'Bannon, {\it {On
  Holographic p-wave Superfluids with Back-reaction}},
  \href{http://xxx.lanl.gov/abs/0912.3515}{{\tt arXiv:0912.3515}}.

\bibitem{Zeng:2009dr}
H.-B. Zeng, Z.-Y. Fan, and H.-S. Zong, {\it {Superconducting Coherence Length
  and Magnetic Penetration Depth of a p-wave Holographic Superconductor}},
  \href{http://xxx.lanl.gov/abs/0912.4928}{{\tt arXiv:0912.4928}}.

\bibitem{Gubser:2008wz}
S.~S. Gubser and F.~D. Rocha, {\it {The gravity dual to a quantum critical
  point with spontaneous symmetry breaking}},  {\em Phys. Rev. Lett.} {\bf 102}
  (2009) 061601, [\href{http://xxx.lanl.gov/abs/0807.1737}{{\tt
  arXiv:0807.1737}}].

\bibitem{Yarom:2009uq}
A.~Yarom, {\it {Fourth sound of holographic superfluids}},  {\em JHEP} {\bf 07}
  (2009) 070, [\href{http://xxx.lanl.gov/abs/0903.1353}{{\tt
  arXiv:0903.1353}}].

\bibitem{Savvidy:1982wx}
G.~K. Savvidy, {\it Yang-mills classical mechanics as a kolmogorov k-system},
  {\em Phys. Lett.} {\bf B130} (1983) 303.

\bibitem{Son:2002sd}
D.~T. Son and A.~O. Starinets, {\it {Minkowski-space correlators in AdS/CFT
  correspondence: Recipe and applications}},  {\em JHEP} {\bf 09} (2002) 042,
  [\href{http://xxx.lanl.gov/abs/hep-th/0205051}{{\tt hep-th/0205051}}].

\bibitem{Herzog:2002pc}
C.~P. Herzog and D.~T. Son, {\it {Schwinger-Keldysh propagators from AdS/CFT
  correspondence}},  {\em JHEP} {\bf 03} (2003) 046,
  [\href{http://xxx.lanl.gov/abs/hep-th/0212072}{{\tt hep-th/0212072}}].

\bibitem{Gubser:2008sz}
S.~S. Gubser, S.~S. Pufu, and F.~D. Rocha, {\it {Bulk viscosity of strongly
  coupled plasmas with holographic duals}},  {\em JHEP} {\bf 08} (2008) 085,
  [\href{http://xxx.lanl.gov/abs/0806.0407}{{\tt arXiv:0806.0407}}].

\bibitem{Iqbal:2009fd}
N.~Iqbal and H.~Liu, {\it {Real-time response in AdS/CFT with application to
  spinors}},  {\em Fortsch. Phys.} {\bf 57} (2009) 367--384,
  [\href{http://xxx.lanl.gov/abs/0903.2596}{{\tt arXiv:0903.2596}}].

\bibitem{Henningson:1998cd}
M.~Henningson and K.~Sfetsos, {\it {Spinors and the AdS/CFT correspondence}},
  {\em Phys. Lett.} {\bf B431} (1998) 63--68,
  [\href{http://xxx.lanl.gov/abs/hep-th/9803251}{{\tt hep-th/9803251}}].

\bibitem{Mueck:1998iz}
W.~Mueck and K.~S. Viswanathan, {\it {Conformal field theory correlators from
  classical field theory on anti-de Sitter space. II: Vector and spinor
  fields}},  {\em Phys. Rev.} {\bf D58} (1998) 106006,
  [\href{http://xxx.lanl.gov/abs/hep-th/9805145}{{\tt hep-th/9805145}}].

\bibitem{Faulkner:2009wj}
T.~Faulkner, H.~Liu, J.~McGreevy, and D.~Vegh, {\it {Emergent quantum
  criticality, Fermi surfaces, and AdS2}},
  \href{http://xxx.lanl.gov/abs/0907.2694}{{\tt arXiv:0907.2694}}.

\bibitem{Gubser:2009qt}
S.~S. Gubser and F.~D. Rocha, {\it {Peculiar properties of a charged dilatonic
  black hole in AdS5}},  \href{http://xxx.lanl.gov/abs/0911.2898}{{\tt
  arXiv:0911.2898}}.

\bibitem{Bloch:1968ab}
F.~Bloch, {\it Flux quantization and dimensionality},  {\em Physical Review}
  {\bf 166} (1968), no.~2 415--423.

\bibitem{Zhang:1997so}
S.-C. Zhang, {\it A unified theory based on $so(5)$ symmetry of
  superconductivity and antiferromagnetism},  {\em Science} {\bf 275} (1997),
  no.~5303 1089--1096.

\bibitem{RevModPhys.76.909}
E.~Demler, W.~Hanke, and S.-C. Zhang, {\it $ so (5) $ theory of
  antiferromagnetism and superconductivity},  {\em Rev. Mod. Phys.} {\bf 76}
  (Nov, 2004) 909--974.

\end{thebibliography}\endgroup

\end{document}